\documentclass[12pt,reqno,a4paper]{amsart}
\usepackage{amsmath,amssymb,dsfont,graphicx,cite,rotating}
 
\usepackage{epsfig}

\def\beq{\begin{equation}}
\def\eeq{\end{equation}}
\def\bea{\begin{eqnarray}}
\def\eea{\end{eqnarray}}
\newtheorem{theorem}{Theorem}

\newcommand{\CV}{{\mathcal V}}
\newcommand{\CC}{{\mathcal C}}

\newcommand{\CE}{{\mathcal E}}

\newcommand{\cnj}{{\mathrm {cn}}}
\newcommand{\dnj}{{\mathrm {dn}}}
\newcommand{\snj}{{\mathrm {sn}}}

\def\ep{\epsilon}

\def\la{\lambda}

\def\tilde{\widetilde}
\def\hat{\widehat}
\def\endpf{\begin{flushright}$\square$\end{flushright}}
\def\alg{{\mathfrak g}}
\def\su2{{\mathfrak {su}}(2)}
\def\e3{{\mathfrak {e}}(3)}

\begin{document}
\title[Three-dimensional discrete systems of Hirota-Kimura type]
{\bf Three-dimensional discrete systems of Hirota-Kimura type and deformed Lie-Poisson algebras}
\author[ANDREW N.W. HONE and MATTEO PETRERA]
{ANDREW N.W. HONE$^{\dag}$ and MATTEO PETRERA$^{\flat}$}

\maketitle

{\footnotesize{

\centerline{\it $^{\dag}$Institute of Mathematics, Statistics and Actuarial Science,}
\centerline{\it University of Kent,}
\centerline{\it Canterbury CT2 7NF, UK}
\vspace{.1truecm}
\centerline{e-mail: \texttt{A.N.W.Hone@kent.ac.uk}}

\vspace{.2truecm}

\centerline{\it $^{\flat}$Dipartimento di Fisica,}
\centerline{\it Universit\`a degli Studi Roma Tre and Sezione INFN, Roma Tre,}
\centerline{\it Via della Vasca Navale 84, 00146 Roma, Italy}
\vspace{.1truecm}
\centerline{e-mail: \texttt{petrera@fis.uniroma3.it}}

}}

\begin{abstract}
Recently Hirota and Kimura presented a new discretization of the 
Euler top with several remarkable properties. In particular this 
discretization shares with the original continuous system 
the feature that it is an algebraically
completely integrable bi-Hamiltonian system in three dimensions. 
The Hirota-Kimura discretization scheme turns out to be 
equivalent to an approach to numerical integration 
of quadratic vector fields that was introduced by Kahan, who applied it  
to the two-dimensional Lotka-Volterra system. 

The Euler top is 
naturally written in terms of the $\mathfrak{so}(3)$ Lie-Poisson algebra. Here we 
consider 
algebraically 
integrable systems 
that are associated with pairs of Lie-Poisson algebras in three dimensions, 
as presented by G\"umral and Nutku, and construct birational maps 
that discretize them according to the scheme of Kahan and Hirota-Kimura. 
We show that the maps thus obtained are also bi-Hamiltonian, with pairs 
of compatible Poisson brackets that are one-parameter deformations 
of the original Lie-Poisson algebras, and hence they are completely 
integrable. For comparison, we also present analogous discretizations  
for 
three bi-Hamiltonian systems that have a transcendental invariant, and finally  
we analyze all of the maps obtained from the viewpoint of 
Halburd's 
Diophantine integrability criterion. 
\end{abstract}

 
\section{Introduction} \label{sec0}

The problem of numerical integration, 
namely that of approximating the flow of a
smooth vector field 
by an iterative scheme given in terms of a difference equation or a map, 
is one of the central problems of numerical analysis. 
If the underlying differential equation is Hamiltonian, or volume-preserving,  
or has some other important geometrical feature (such as being invariant under 
the action of a Lie group of symmetries), then as far as possible one 
would like to select a discretization scheme which preserves this feature, 
and this has led to the development of geometrical integration methods 
\cite{budd}. For the special case of completely integrable 
systems, ideally one would like to obtain discretizations which 
are themselves completely integrable. The area of integrable 
discretization has been developed quite extensively, especially 
from the Hamiltonian viewpoint, and a comprehensive review of 
the field can be found in the monograph \cite{S}. In this paper 
we are concerned with 
a novel approach to 
discretization, which was    
used by Hirota and Kimura to obtain new integrable 
discrete analogues of the Euler and Lagrange tops 
\cite{HK1, HK2}.

The discretization method studied in this paper seems to be
introduced in the geometric integration literature by W. Kahan in
the unpublished notes \cite{K}. It is applicable to any system of
ordinary differential equations for ${\bf x}:\mathbb R\to\mathbb R^n$ with a
quadratic vector field 
\begin{equation}\nonumber
\dot{{\bf x}}=Q({\bf x})+B{\bf x}+{\bf c},
\end{equation}
where each component of $Q:\mathbb R^n\to\mathbb R^n$ is a quadratic form,
while $B\in{\rm Mat}_{n\times n}(\mathbb R )$ 
and ${\bf c}\in\mathbb R^n$. Kahan's
discretization reads as
\begin{equation}\label{eq: Kahan gen}
\frac{\widetilde{{\bf x}}-{\bf x}}{2\epsilon}=Q({\bf x},\widetilde{{\bf x}})+
\frac{1}{2}B({\bf x}+\widetilde{{\bf x}})+{\bf c},
\end{equation}
where
\[
Q({\bf x},\widetilde{{\bf x}})=\frac{1}{2}\left[Q({\bf x}+\widetilde{{\bf x}})-Q({\bf x})-
Q(\widetilde{{\bf x}})\right],
\]
is the symmetric bilinear form corresponding to the quadratic form
$Q$. Here and below we use the following notational convention
which will allow us to omit a lot of indices: for a sequence
${\bf x}:\mathbb Z\to\mathbb R$ we write ${\bf x}$ for ${\bf x}_k$ and $\widetilde{{\bf x}}$ for
${\bf x}_{k+1}$. Eq. (\ref{eq: Kahan gen}) is {linear} with respect
to $\widetilde {\bf x}$ and therefore defines a {rational} map
$\widetilde{{\bf x}}=f({\bf x},\epsilon)$. Clearly, this map approximates the
time-$(2\epsilon)$-shift along the solutions of the original
differential system, so that ${\bf x}_k\approx {\bf x}(2k\epsilon)$. (We have
chosen a slightly unusual notation $2\epsilon$ for the time step,
in order to avoid appearance of powers of 2 in numerous
formulae; a more standard choice would lead to changing
$\epsilon\mapsto\epsilon/2$ everywhere.) Since Eq. (\ref{eq: Kahan
gen}) remains invariant under the interchange
${\bf x}\leftrightarrow\widetilde{{\bf x}}$ with the simultaneous sign
inversion $\epsilon\mapsto-\epsilon$, one has the {
reversibility} property $
f^{-1}({\bf x},\epsilon)=f({\bf x},-\epsilon).
$
In particular, the map $f$ is {birational}.

Kahan applied this discretization scheme to the famous
Lotka-Volterra system and showed that in this case it possesses a
very remarkable non-spiralling property. Some further applications of
this discretization have been explored in \cite{KHLI,roeger}.

The next, even more intriguing, appearance of this discretization
was in the two papers by R. Hirota and K. Kimura who (being
apparently unaware of the work by Kahan) applied it to two famous
{\it integrable} systems of classical mechanics, the Euler top and
the Lagrange top \cite{HK1, HK2}. Surprisingly, the Kahan-Hirota-Kimura
discretization scheme produced {\it integrable  maps} 
in both the Euler and the Lagrange
cases of rigid body motion. 
Even more
surprisingly, the mechanism which assures integrability in these
two cases seems to be rather different from the majority of
examples known in the area of integrable discretizations, and,
more generally, integrable maps, cf. \cite{S}. We shall use the term ``Hirota-Kimura type
discretization'' for 
Kahan's discretization in the context of
integrable systems. 

In the recent paper \cite{PPS} the Hirota-Kimura
integrability mechanism has been 
further investigated 
and its application to the integrable
(six-dimensional) Clebsch system has been considered. 
The integrability of the Hirota-Kimura type
discretization of the Clebsch system 
has been established, in the sense of:
{\it i}) existence, for every initial point, of a four-dimensional pencil of quadrics
containing the orbit of this point; {\it ii}) existence of four functionally independent integrals of
motion. Note that for the purposes of paper  \cite{PPS}, integrability of a dynamical system is synonymous with the
existence of a sufficient number of functionally independent
conserved quantities, or integrals of motion, that is, functions
constant along the orbits. Other aspects of the notion of integrability, 
such as Hamiltonian properties or
explicit solutions, still require further investigation. However, 
it is known that algebraically completely integrable 
cases of geodesic flow on $SO(4)$ are related  
to the intersection of four quadrics in $\mathbb{P}^6$ \cite{adlervanm}.  
The Hirota-Kimura method of discretization
has been recently applied to the
classical three-dimensional nonholonomic Suslov problem in \cite{DG}.

The above examples of Hirota-Kimura type discretizations suggested the 
following 

\vspace{.1truecm}

\noindent 
{\bf{Conjecture}} \cite{PPS}.   
{\it For any algebraically completely integrable system with a
quadratic vector field, its Hirota-Kimura type discretization is 
algebraically completely integrable.}

\vspace{.1truecm}

\noindent
Since algebraically completely integrable systems 
generically correspond to linear flows on abelian varieties 
\cite{vanhaecke}, 
this statement should be related to addition theorems for
multi-dimensional theta-functions. 

The aim of this paper is both 
to study how  
this novel method of discretization applies to a set of 
algebraically 
integrable systems in three dimensions, and to see how these 
results compare with the analogous discretizations of 
some quadratic vector fields with transcendental invariants. 
The former set of systems considered are 
\textit{algebraically integrable} in the sense that they 
have a sufficient number of algebraic integrals in 
involution; however, there are various other (more stringent)  
notions of algebraic complete integrability.  
Through this study we are able both to verify the above 
conjecture for a new set of examples, and to gain further 
understanding of how the integrability of 
the discretization depends on the algebraic nature (or otherwise) of the 
integrals of motion in the original continuous system. 
Kahan's discrete Lotka-Volterra system illustrates the 
subtlety of this dependence, as we now describe. 

Kahan used his approach to discretize the Lotka-Volterra system 
$$ 
\dot{x} = x(1-y) , \qquad \dot{y}=y(x-1), 
$$ 
which preserves the Poisson bracket $\{ x,y\}=xy$, or equivalently, 
the symplectic form $1/(xy)\, dx\wedge dy$. 
This is an integrable system with one degree of freedom, 
$\dot x = \{ x,H\}$, $\dot y = \{ y,H\}$  with Hamiltonian  
$$ 
H=\log xy -(x+y). 
$$   
Kahan's discretization for this system reads \cite{K} 
\beq
\label{kahandlv} 
\frac{\tilde x - x }{\epsilon} = \tilde x + x - \tilde x y - x\tilde y, 
\qquad 
\frac{\tilde y - y }{\epsilon} = -\tilde y - y + \tilde x y + x\tilde y.   
\eeq 
This discretization preserves the same symplectic structure 
as the original system of  
ordinary differential equations, for which it 
provides a numerically stable integration scheme which appears to retain 
the qualitative features of the continuous orbits (which are closed 
curves $H=\,$constant in the positive quadrant $x>0$, $y>0$) \cite{ss}. 

In fact, as noted in \cite{ps07preprint}, Kahan's  
discrete Lotka-Volterra system is algebraically integrable for 
$\epsilon =\pm 1$. To be precise, 
when $\epsilon =1$, 
it reduces 
to the second order recurrence $x_{n+1}x_{n-1}=x_n(2-x_n)$ for the 
$x$ coordinate, which belongs to the 
class of antisymmetric QRT maps studied 
in \cite{wgr}. 
This recurrence is linearizable (the iterates 
satisfy a linear recurrence of sixth order \cite{hone2}),  
and the map has the 
integral 
$$ 
\hat H = \frac{x^2}{y^2}+\frac{y^2}{x^2} 
+4\left(\frac{1}{x}-\frac{1}{y}\right)^2(1-x-y), 
$$ 
which (for fixed $\hat H$) 
defines a quartic curve of genus zero; this is also 
an integral for 
$\epsilon =-1$ (which can be seen immediately 
from the reversibility 
property of Kahan's discretization scheme). 
However, 
for other non-zero values of $\epsilon$, Kahan's  discrete 
Lotka-Volterra system should not be algebraically integrable;  we 
present some numerical evidence for this in section 7 below.  
Indeed, since the 
integral $H$ for the original system is transcendental, 
from continuity arguments
one would expect that (at least for small enough $\epsilon$) 
any integral of the 
the discretization should be transcendental as well.  
Further numerical studies, as mentioned in \cite{PPS}, indicate that 
this discrete system may well be non-integrable, with characteristics 
of chaos only evident by zooming in deeply on regions of the phase plane.     

The outline of the paper is as follows. In the next section, we briefly 
review the Euler top together with the discrete Euler top found by 
Hirota and Kimura. 
In section 3 we describe six quadratic bi-Hamiltonian flows in three 
dimensions, which were presented in \cite{nutku} (extending 
results in \cite{bw}), 
and are associated with pairs of real three-dimensional 
Lie 
algebras. Moreover,  each of these systems, which we denote by 
$\CE_i$ for $i=1,\ldots ,6$, is algebraically 
integrable; 
the system $\CE_6$ is equivalent to a special case of the Euler top.   
The fourth section is devoted to applying the Hirota-Kimura 
discretization scheme to these six systems, to obtain 
discrete systems (or maps) 
in three dimensions which we denote by $d\CE_i$, 
and in section 5 we 
present the explicit solutions of these maps for $i=1,\ldots ,5$ 
(the case $i=6$ being already included in the work of Hirota and Kimura 
\cite{HK1}). Section 6 is concerned with applying the same 
discretization method to three 
other 
bi-Hamiltonian systems from \cite{nutku} which have transcendental 
integrals. In section 7 we present the results of applying Halburd's   
Diophantine integrability test to each of the maps obtained, 
and prove that all but one of them 
are Diophantine integrable in the sense 
of \cite{halburd}. The final section is devoted to some conclusions.

\section{Euler top and its Hirota-Kimura type discretization} \label{sec1}

The $\mathfrak{so}(3)$ Euler top is a well-known three-dimensional bi-Hamiltonian system belonging
to the realm of classical mechanics \cite{RSTS}.
The differential equations of motion of the Euler top read
\begin{equation}\label{eq: ET x}
\dot{x}=\alpha_1 y z,\qquad \dot{y}=\alpha_2 z x, \qquad
\dot{z}=\alpha_3 xy,
\end{equation}
with $\alpha_i$ being real parameters of the system. We recall  that this
system can be explicitly integrated in terms of elliptic
functions, and admits two functionally independent integrals of
motion. Indeed, a quadratic function
$H({\bf x})=\gamma_1x^2+\gamma_2y^2+\gamma_3z^2$ is an integral
for Eqs. (\ref{eq: ET x}), if $\gamma_1\alpha_1+\gamma_2\alpha_2+\gamma_2\alpha_2=0$.
In particular, the following three functions are integrals of
motion:
\begin{equation}\nonumber
H_1=\alpha_3y^2-\alpha_2z^2,\qquad
H_2=\alpha_1z^2-\alpha_3x^2,\qquad
H_3=\alpha_2x^2-\alpha_1y^2.
\end{equation}
Clearly, only two of them are functionally independent because of
$\alpha_1H_1+\alpha_2H_2+\alpha_3H_3=0$.

The Hirota-Kimura discretization of the Euler top
introduced in \cite{HK1} reads as
\begin{equation}\label{eq: dET x}
\renewcommand{\arraystretch}{1.3}
\left\{\begin{array}{l}
\widetilde{x}-x=\epsilon\alpha_1(\widetilde{y}z+y\widetilde{z}),\\
\widetilde{y}-y=\epsilon\alpha_2(\widetilde{z}x+z\widetilde{x}),\\
\widetilde{z}-z=\epsilon\alpha_3(\widetilde{x}y+x\widetilde{y}).
\end{array}\right.
\end{equation}
Thus, the map $f:{\bf x}\mapsto\widetilde{\bf x}$ obtained by solving
(\ref{eq: dET x}) for $\widetilde{{\bf x}}$, is given by:
\begin{equation}\label{eq: dET map}
\widetilde{{\bf x}} =f({\bf x},\epsilon)=A^{-1}({\bf x},\epsilon){\bf x}, \qquad
A({\bf x},\epsilon)=
\begin{pmatrix}
1 & -\epsilon\alpha_1 z & -\epsilon\alpha_1 y \\
-\epsilon\alpha_2 z & 1 & -\epsilon\alpha_2 x \\
-\epsilon\alpha_3 y & -\epsilon\alpha_3 x & 1
\end{pmatrix} .
\end{equation}

Apart from the Lax representation which is still unknown, the
discretization (\ref{eq: dET map}) exhibits all the usual features
of an integrable map: an invariant volume form, a bi-Hamiltonian
structure  (that is, two compatible invariant Poisson structures),
two functionally independent conserved quantities in involution,
and solutions in terms of elliptic functions. For further
details about the properties of this discretization we refer to
\cite{HK1} and \cite{PS}.

\section{Some bi-Hamiltonian flows related to real three-dimensional Lie algebras} \label{sec2}

Hamiltonian systems in three dimensions provide the simplest non-trivial 
examples  
of degenerate Poisson structures, where the rank of the Poisson tensor 
is less than the dimension of the phase space. In three dimensions, 
a non-trivial Poisson tensor $P$ 
has rank two at generic points of the phase 
space, which means that (at least locally) 
there exists a Casimir function $K$ 
and another function $\phi$ such that
\beq \label{pb3d} 
\{x_j,x_k\}= \varepsilon_{jkl}\, \phi\,  
\frac{\partial K}{\partial x_l}
\eeq 
in local coordinates $x_1,x_2,x_3$; cf. Theorem 5 in \cite{gmp}. 
This can be expressed in invariant form, by 
using the standard volume three-form 
$\Omega =dx_1\wedge dx_2\wedge dx_3$ 
in $\mathbb{R}^3$ to associate $P$ with the one-form  
$J=P \lrcorner\, \Omega =\phi\, dK$. An important thing to observe 
from the form of the Poisson bracket (\ref{pb3d}) is that 
in three dimensions the Poisson tensor can be multiplied by an arbitrary 
function while preserving the Jacobi identity. 

Given an Hamiltonian system  
$$ 
\dot {\bf x} = \{ {\bf x}, H\}
$$ 
defined in terms of the bracket (\ref{pb3d}) with 
an Hamiltonian function $H$ (functionally independent of $K$), 
it is clear that the equations of motion have two 
independent integrals, namely $H$ and $K$. Moreover, by fixing the 
value of the Casimir function (which may not be defined 
everywhere), we can regard this locally as a system 
with one degree of freedom which is integrable 
on each of the two-dimensional symplectic leaves $K=\mathrm{constant}$. 
However, for complete integrability the global 
existence of $H$ and 
$K$ is required. 

G\"umral and Nutku made a detailed study of the geometry of 
three-dimensional Poisson 
structures, and considered the conditions for the existence of 
globally integrable bi-Hamiltonian structures \cite{nutku}. 
For a given three-dimensional system to be 
bi-Hamiltonian it
is
necessary and sufficient that the Jacobian at 
an arbitrary point be a Poisson 
tensor and that there exist two globally
defined and (almost everywhere) functionally independent integrals of motion.
Associated with two independent integrals   
$K$ and $H$, there are two compatible 
Poisson tensors, such that $K$ is the Casimir for one   
Poisson structure while $H$ is the Casimir for the other. 
In other words, if the dimension is three then  
two compatible Poisson tensors are completely determined 
by the constants of motion, 
and according to a relevant
Theorem by Magri \cite{MM},  
provided certain technical conditions are satisfied,  
bi-Hamiltonian systems are
completely integrable in the sense of Liouville-Arnold.
Furthermore, in this 
setting there is an invariant volume form $\Omega$ 
(not necessarily canonical) which is preserved by the bi-Hamiltonian 
flow. An important example of such flows 
corresponds to Nambu 
mechanics \cite{nambu}, given by 
$$ 
\dot {\bf x} =\nabla H \times \nabla K,  
$$    
which in these coordinates gives a divergenceless vector field 
($\mathrm{div}\, \dot {\bf x}=0$); this means that the canonical 
measure is preserved by the flow. In particular, the Euler top 
is an example of Nambu mechanics in three dimensions; for other examples 
of Nambu mechanics in optics and elsewhere, see \cite{holm}.

In \cite{nutku} the authors present a list of all non-trivial 
bi-Hamiltonian flows that are associated with pairs of real 
three-dimensional Lie algebras and their Casimir invariants 
(as described in \cite{w}); this list extends results in \cite{bw}. 
To be more precise, they consider pairs of real Lie-Poisson algebras 
defined by pairs of linear Poisson structures $P,Q$, and write down 
vector fields $ \dot {\bf x}= {\ V}$ satisfying 
$$ 
{\ V}=-P\, dK =-\frac{1}{c} \, Q\, dH ,
$$ 
where $K$ is the Casimir for $Q$, while 
$H$ is the Casimir for $P$  
(and minus signs are included in order to be consistent 
with the conventions of G\"umral and Nutku). 
In \cite{nutku} twelve such systems are presented, extending a list 
in \cite{bw}, and each flow preserves a corresponding 
measure given in coordinates $(x_1,x_2,x_3)=(x,y,z)$ 
by 
$$ 
\Omega =c\, dx\wedge dy\wedge dz, 
$$ 
related to the standard volume form by the conformal factor (or multiplier) 
$c$.\footnote{In fact, on page 5704 of \cite{nutku} the authors state that 
the given systems are all ``with multiplier unity", and denoting 
the multiplier by $M$ they say ``these equations have $M=1$ [...] they 
are Nambu mechanics representatives", but as should be clear from 
Table 1 this is 
not the case: two of the systems given there have a non-constant 
multiplier. For systems with non-constant multiplier, it is     
remarked in \cite{nutku} that they may be only locally 
(but not globally) equivalent 
to Nambu mechanics, by a suitable change of coordinates.}
    
To begin with, 
we shall be concerned with only six out of the twelve systems 
in G\"umral and Nutku's list, namely 
the ones which have non-transcendental integrals of motion. 
They read
\bea
&& 
\CE_1: \qquad \dot x = -x^2, 
\; \; \, \; \; \; \; \, \; \; \; \;  \; \;  \;  \qquad 
\dot y = - x y, 
\qquad \qquad \; \; 
\dot z = 2 y^2 +x z \, ; \label{3a}\\
&& \CE_2: \qquad \dot x = -x^2, 
\; \; \, \; \; \; \; \, \; \; \; \;  \; \;  \; \qquad 
\dot y = x y, 
\qquad \qquad \; \; \; \; \; 
\dot z = -2  y^2 +x z \, ; \label{3b}\\
&& 
\CE_3: \qquad \dot x = -xz , 
\qquad \qquad \! \qquad 
\dot y = -  y z,
 \qquad \qquad \, \; \; 
 \dot z =  x^2+ y^2 \, ; \label{4a}\\
 &&\CE_4: \qquad  \dot x = -xz , 
\qquad \qquad \! \qquad 
\dot y =  y z,
 \qquad \qquad \; \; \; \, \; \; 
 \dot z =  x^2 - y^2 \, ; \label{4b}\\
 &&  
\CE_5: \qquad  \dot x = xy , 
 \; \; \; \;  \; \,  \;\; \qquad \; \; \qquad 
 \dot y = - x^2, 
 \qquad \; \, \; \,\, \; \; \; \; \; \; 
 \dot z = y(2x -z) \, ; \label{6} \\
&& 
\CE_6: \qquad \dot x = y (2z -x), \qquad 
 \; \; \; \; \; \; 
 \dot y = x^2 - z^2, 
 \qquad \; \; \; \; \; 
 \dot z = y(z -2 x) \, .\label{8}
\eea
These six systems are examples of algebraically completely 
integrable systems, in the sense that in each case the 
integrals are algebraic (in fact, rational) functions of the 
coordinates $x,y,z$ for which the Poisson structures $P,Q$ are linear. 
(We consider two other examples where there is
one transcendental invariant in section 6.)  
 
The corresponding linear 
Poisson structures and integrals of motion are given in Table \ref{tab1}. For instance, the flow
$\CE_1$, given in Eq. (\ref{3a}), admits the bi-Hamiltonian structure given by the compatible pair
$( P^{(1)} ,c_1^{-1}Q^{(1)})$, where
\bea
&& P^{(1)}:\qquad P_{12}^{(1)} = \{x,y \}=0, \qquad P_{23}^{(1)} = \{y,z \}=y, \qquad P_{31}^{(1)} = \{z,x \}=-x, \nonumber \\
&& Q^{(1)}:  \qquad Q_{12}^{(1)} = \{x,y \}=x,  
\qquad Q_{23}^{(1)} = \{y,z \}=z, \qquad Q_{31}^{(1)} = \{z,x \}= 2y, \nonumber 
\eea
with conformal factor $c_1=1/x^2$.
The quantities $H_1= y/x$ and $K_1=zx +y^2$, preserved by the flow,
are respectively the Casimir functions of $P^{(1)}$ and $Q^{(1)}$. This is equivalent to say that the following
Lenard-Magri chain \cite{MM} is satisfied:
$$
P^{(1)} d H_1=0, \qquad
P^{(1)} d K_1 = \frac{1}{c_1} Q^{(1)} d H_1= - (-x^2, - x y,  2 y^2 +x z)^T \qquad 
Q^{(1)} d K_1=0,
$$
where $dH_1$ and $dK_1$ denote the differentials of the functions $H_1$ and $K_1$ respectively. The same scheme holds
for the flows $\CE_i$ with $2 \leq i \leq 6$. 

In Table \ref{tab1} 
there are actually just five independent Lie-Poisson structures, namely
$P^{(1)}=P^{(3)}, P^{(2)}=P^{(4)}, P^{(5)}, P^{(6)}=Q^{(1)}=Q^{(2)}=Q^{(5)}, Q^{(3)}=Q^{(4)}=Q^{(6)}$,
corresponding respectively to the Casimir functions $H_1=H_3, H_2=H_4, H_5, H_6=K_1=K_2=K_5, K_3=K_4=K_6$.
The last column in Table  \ref{tab1} 
gives the associated real three-dimensional Lie algebras; see \cite{w} for more details.
Observe that 
the flows $\CE_4$ and  $\CE_6$ each 
correspond to a particular case of the equations of motion  (\ref{eq: ET x})
of the $\mathfrak{so}(3)$ Euler top. 
More precisely, for $\CE_4$ one has to make the change of variables 
$(x,y,z) \mapsto (x-y,x+y,z)$ and fix the parameters 
so that $(\alpha_1,\alpha_2,\alpha_3)=(-1,-1,1)$, while 
for $\CE_6$ one takes  
$(x,y,z) \mapsto (x-z,x+z,y)$ and the parameters 
are $(\alpha_1,\alpha_2,\alpha_3)=(1,-3,1)$.

{\footnotesize{
\begin{table}[h!]\centering\caption{Lie-Poisson structures, invariants, conformal factors and
related real three-dimensional Lie algebras} \label{tab1}
\begin{tabular}{|l||c|c|c||c|c|c|| c|c||c|| c| }
 \hline
 $i$ & $P_{12}^{(i)}$ & $P_{23}^{(i)}$ & $P_{31}^{(i)}$ & $Q_{12}^{(i)}$ & $Q_{23}^{(i)}$ & $Q_{31}^{(i)}$  & $H_i$ & $K_i$ &$ c_i$ 
 & $\alg$ \\ \hline \hline
 & & && & &&&  &&\\
$1$ & $0$  &  $y$& $-x $ & $x $ & $z$ & $ 2 y$ & {$\displaystyle \frac{y}{x} $}& $zx +y^2$ &  {$\displaystyle \frac{1}{x^2} $} 
& $A_{3,3}$, $\mathfrak{sl}(2, \mathbb R)$ \\
 & & &&&&& &&&\\
$2$ & $0$ & $ -y$ & $-x  $ & $x $& $z$ & $2 y$&  {$\displaystyle xy$}& $zx +y^2$ & $1$ & 
$\mathfrak e (1,1)$, $\mathfrak{sl}(2, \mathbb R)$\\
 & & && &&&&& &\\
$3$& $0$   & $y $ & $-x$ & $z$  & $x$&$y$ &  {$\displaystyle \frac{y}{x} $}& $\frac12 (x^2 +y^2+z^2)$ &  {$\displaystyle \frac{1}{x^2} $}
& $A_{3,3}$, $\mathfrak{so}(3)$\\
 & & && &&&&&&  \\
$4$&  $0$ &  $-y$& $-x $  & $z $& $x$& $y $ &  {$\displaystyle xy $}& $\frac12 (x^2 +y^2+z^2)$ &  $1$& 
$\mathfrak e (1,1)$, $\mathfrak{so}(3)$\\
 & & && &&&&&&  \\
$5$& $0$ & $x $& $y$ & $x $& $z $ & $2 y $ &  $\frac12 (x^2 +y^2)$ & $zx +y^2$ & $-1$& 
$\mathfrak e (2)$, $\mathfrak{sl}(2, \mathbb R)$\\
 & & && &&&&& & \\
$6$ & $x $& $z $& 
$2 y $ &  $z  $ & $x $& 
$y $ & $zx+y^2$  &  $\frac12 (x^2 +y^2+z^2)$ & $-1$ & $\mathfrak{sl}(2, \mathbb R)$, $\mathfrak{so}(3)$ \\
 & & && &&&& &&  \\ \hline
\end{tabular}
\end{table}
}}

\section{Hirota-Kimura type discretization of the flows $\CE_i$} \label{sec3}

The goal of this section is to show that 
Hirota-Kimura type discretizations of the bi-Hamiltonian flows
$\CE_i$, $1 \leq i \leq 6$, provide completely integrable discrete-time 
systems. The following 
result holds.

\begin{theorem} 

The Hirota-Kimura type discretizations of the bilinear flows $\CE_i$, $1 \leq i \leq 6$, given in Eqs. (\ref{3a}-\ref{8}), read
\beq
d \CE_i: \qquad  \tilde {\bf{x}}= A_i^{-1}({\bf{x}}; \ep)\, 
 {\bf{x}}= A_i ( \tilde {\bf{x}}; -\ep)\, {\bf{x}}, \label{df56}
\eeq
where the matrices $A_i({\bf{x}}; \ep)$ are given in Table \ref{tab2}.
The quantities $H_i(\ep), K_i(\ep)$, given in Table \ref{tab2}, are integrals of motion for the maps (\ref{df56}).
Moreover the maps (\ref{df56}) preserve the volume 
form
\beq
\Omega_i = \frac{c_i}{H_i K_i } dx \wedge dy \wedge dz, \qquad 1 \leq i \leq 6. \label{form}
\eeq
\end{theorem}

\begin{table}[h!]\centering\caption{ Matrices $A_i({\bf{x}}; \ep)$ and discrete integrals of motion} \label{tab2}
\begin{tabular}{|l||c||c|c|}
 \hline
 $i$ & $A_i({\bf{x}};\ep)$ & $H_i(\ep)$ & $K_i(\ep)$ \\ \hline \hline
 & & &  \\
$1$ & $\begin{pmatrix} 1 +2\ep x &0&0 \\
\ep y & 1 +\ep  x  &0 \\
-\ep z  & -4 \ep  y& 1- \ep x
\end{pmatrix}$ & $H_1$ & ${\displaystyle \frac{K_1}{1-\ep^2x^2}}$ \\
 & & &  \\
$2$ & $\begin{pmatrix} 1 +2\ep x &0&0 \\
-\ep  y & 1 -\ep  x  &0 \\
-\ep z  & 4 \ep y& 1- \ep x
\end{pmatrix}$  & ${\displaystyle \frac{H_2}{1-\ep^2x^2}}$ & ${\displaystyle \frac{K_2}{1-\ep^2x^2}}$ \\
 & & &  \\
$3$& $\begin{pmatrix} 1 +\ep z&0&\ep x \\
0& 1 +\ep  z  & \ep  y   \\
-2\ep x  &  -2\ep  y& 1
\end{pmatrix}$ & $H_3$ & ${\displaystyle \frac{K_3}{1+\ep^2(x^2+y^2)}}$ \\
 & & &  \\
$4$& $\begin{pmatrix} 1 +\ep z&0&\ep x \\
0& 1 -\ep z  &- \ep  y   \\
-2\ep x  &  2\ep  y& 1
\end{pmatrix}$ & ${\displaystyle \frac{H_4}{1+\ep^2(x^2+y^2)}}$ & ${\displaystyle \frac{K_4}{1+\ep^2(x^2+y^2)}}$  \\
 & & &  \\
$5$& $\begin{pmatrix} 1 -\ep y &- \ep x &0 \\
2\ep x & 1 &0 \\
- 2\ep y & -\ep (2x -z ) & 1+\ep y
\end{pmatrix}$ & ${\displaystyle \frac{H_5}{1+\ep^2x^2}}$ & ${\displaystyle \frac{K_5}{1+\ep^2x^2}}$ \\
 & & & \\
$6$ & $\begin{pmatrix} 1 +\ep y & \ep (x -2z)& -2\ep y \\
-2\ep x & 1 & 2\ep z \\
2\ep y  & \ep (2x-z ) & 1- \ep y
\end{pmatrix}$ & ${\displaystyle \frac{H_6}{1-{3} \ep^2 x z}}$ & ${\displaystyle \frac{K_6}{1-{3}\ep^2 x z}}$ \\
 & & &  \\ \hline
\end{tabular}
\end{table}

\noindent 
Note that for small $\ep$    
the birational maps (\ref{df56}) approximate
the time shift along the trajectories of the 
corresponding  
continuous equations of motion (\ref{3a}-\ref{8}). The same  
invariant volume form (\ref{form}), which is 
independent of $\ep$, is preserved by both the  
continuous and the discrete systems.   

{\bf Proof:} We shall prove Theorem 1 for just one case, 
namely $i=5$. The remaining cases can be proved by similar straightforward
computations.

The Hirota-Kimura discretization of the flow $\CE_5$, 
given by Eq. (\ref{6}), reads explicitly as
\begin{equation}\label{st}
\renewcommand{\arraystretch}{1.3}
\left\{\begin{array}{l}
\widetilde{x}-x=  \ep (x \tilde y + \tilde x y),\\
\widetilde{y}-y=  - 2\ep x \tilde x,\\
\widetilde{z}-z=  \ep(2 x \tilde y + 2  \tilde x y - y \tilde z - \tilde y z), 
\end{array}\right.
\end{equation}
that is 
$$
\tilde {\bf{x}}= A_5^{-1}({\bf{x}}; \ep) {\bf{x}}= A_5 ( \tilde {\bf{x}}; -\ep) {\bf{x}},
$$
with
$$
 A_5({\bf{x}}; \ep)=\begin{pmatrix} 1 -\ep y &- \ep x &0 \\
2\ep x & 1 &0 \\
- 2\ep y & -\ep (2x -z ) & 1+\ep y
\end{pmatrix}.
$$
The fact that the quantities
\beq
H_5(\ep)= \frac12 \frac{x^2 +y^2}{1+\ep^2x^2},
\qquad K_5(\ep)=  \frac{zx +y^2}{1+\ep^2x^2}, 
\nonumber
\eeq
are integrals of motion of the map (\ref{st}) is proved by the following computation. Equation
$\tilde H_5(\ep) = H_5(\ep)$ means that
$$
(\tilde x - x)(\tilde x + x) + (\tilde y - y )(\tilde y + y) = -\ep^2( x \tilde y - \tilde x y )(x \tilde y + \tilde x y),
$$
that is, using Eq. (\ref{st}), 
$$
\frac12 (x \tilde y + \tilde x y)(\tilde x + x) - x \tilde x (\tilde y + y) = - \frac 12 (x \tilde y - \tilde x y) (\tilde x - x),
$$
which is an algebraic identity. A similar computation shows that equation $\tilde K_5(\ep) = K_5(\ep)$ is identically
satisfied.

We now prove that the map (\ref{st}) preserves the volume 
form
$$
\Omega_5 = - \frac{2}{(xz+y^2 )( x^2+y^2) } dx \wedge dy \wedge dz,
$$
which is equivalent to saying that
$$
\det \frac{\partial \tilde {\bf x}}{ \partial {\bf x}} = \frac{(\tilde x \tilde z+ \tilde y^2 )( \tilde x^2+\tilde y^2) }{(xz+y^2 )( x^2+y^2) }.
$$
First of all we note that differentiating Eq. (\ref{st}) with respect to $x,y,z$ one obtains the columns of the matrix equation
$$
A_5({\bf{x}}; \ep) \, \frac{\partial \tilde {\bf x}}{ \partial {\bf x}} = A_5 ( \tilde {\bf{x}}; -\ep) .
$$
Computing determinants lead to
$$
\det \frac{\partial \tilde {\bf x}} { \partial {\bf x}} = \frac{ \det A_5 ( \tilde {\bf{x}}; -\ep)}{ \det A_5({\bf{x}}; \ep)}=
\frac{\left( 1 -\ep \tilde y\right) \left( 1+ \ep \tilde y + 2 \ep^2 \tilde x^2 \right)}{
\left( 1 +\ep  y\right) \left( 1- \ep  y + 2\ep^2 x^2 \right)}.
$$
Now, by using the map (\ref{st}), a straightforward computation shows that the 
relation
$$
 \frac{(\tilde x \tilde z+ \tilde y^2 )( \tilde x^2+\tilde y^2) }{(xz+y^2 )( x^2+y^2) }=
 \frac{\left( 1 -\ep \tilde y\right) \left( 1+ \ep \tilde y + 2 \ep^2 \tilde x^2 \right)}{
\left( 1 +\ep  y\right) \left( 1- \ep  y + 2 \ep^2  x^2 \right)}
$$
holds identically. 
\endpf

In the construction of an  invariant Poisson structure for the maps  
(\ref{df56}) we shall make  
use of results 
from \cite{BHQ} (Proposition 15 and
Corollary 16 there), which we restate here. 
Suppose that  $f:M\to M$ is a smooth mapping of an
$n$-dimensional manifold $M$, with an invariant volume
form $\Omega$ (that is, $f^*\Omega = \Omega$). 
Define $\omega$ to be the dual $n$-vector
field to $\Omega$ such that $\omega\lrcorner\,\Omega=1$, 
where as usual the symbol $\lrcorner$ denotes the contraction between
multivector fields and forms. 
It follows that if
$I_1,\ldots,I_{n-2}$ are integrals of $f$ with
$dI_1\wedge\cdots\wedge I_{n-2}\neq 0$, then the bivector field
$\sigma = \omega\lrcorner\, dI_1\cdots\lrcorner\, dI_{n-2}$ is an
invariant Poisson structure for $f$. If $J_1,\ldots,J_{n-2}$ is
another set of independent integrals and $\tau =\omega\lrcorner\,
dJ_1\cdots\lrcorner\, dJ_{n-2}$ is the corresponding Poisson
structure, then $\sigma$ and $\tau$ are compatible, i.e. for any
constants $a$, $b$, the bivector field $a\sigma+b\tau$ is again a
Poisson structure.

In particular for $n=3$, 
if a three-form $\Omega$, given by Eq.  (\ref{form}) in our case, is
invariant under a map $f$ defined by (\ref{df56}), we can define the dual 
trivector field
\[
\omega = \phi(x,y,z)\frac{\partial}{\partial x}\wedge
 \frac{\partial}{\partial y}\wedge
 \frac{\partial}{\partial z}\,,
\]
so that for any integral $I$ of $f$ the bivector field
\begin{equation}
\sigma = \omega  \lrcorner d I =\phi(x,y,z)\left(\frac{\partial I}{\partial
z}\frac{\partial}{\partial x}\wedge\frac{\partial}{\partial
y}+ \frac{\partial I}{\partial x}\frac{\partial}{\partial
y}\wedge\frac{\partial}{\partial z}+\frac{\partial I}{\partial
y}\frac{\partial}{\partial z}\wedge\frac{\partial}{\partial
x}\right)  \nonumber
\end{equation}
is an invariant Poisson structure for $f$, as well as any linear
combination of such bivector fields. Explicitly, the Poisson brackets of
coordinate functions are given by
\begin{equation}
\left\{x,y\right\}= \phi(x,y,z)\frac{\partial I}{\partial z}, \qquad
\left\{y,z\right\}= \phi(x,y,z)\frac{\partial I}{\partial x} , \qquad
\left\{z,x\right\}= \phi(x,y,z)\frac{\partial I}{\partial y}. \nonumber
\end{equation}
Note that the 
inverse 
volume density $\phi(x,y,z)$ can be multiplied by an arbitrary integral of $f$ without violating the Poisson property.

For the maps (\ref{df56}), the invariant Poisson structures $P^{(i)}$, 
$c_i^{-1}Q^{(i)}$ can be computed according to the following formulae:
\beq
P^{(i)}_{jk}(\ep)= -\frac{1}{c_i}\,\frac{H_i K_i}{H_i(\ep) K_i (\ep)}  
\,\varepsilon_{jkl}\,\frac{\partial H_i (\ep)}{ \partial x_l}, \label{p1}  
\eeq 
and
\beq
\frac{1}{c_i}\,Q^{(i)}_{jk}(\ep) 
= \frac{1}{c_i}\,\frac{H_i K_i}{H_i(\ep) K_i (\ep)} 
\,\varepsilon_{jkl}\,\frac{\partial K_i (\ep)}{ \partial x_l}, 
\label{q3}
\eeq
with $ 1 \leq i \leq 6$, the summation convention is assumed 
for the index $l$, and above we have used $(x_1,x_2,x_3)$ 
to denote $(x,y,z)$. (The reader should 
note that these indices $1,2,3$ for the coordinates 
in $\mathbb{R}^3$ should not be confused with the index $n$ used 
to denote iterates of maps in subsequent sections.)
This corresponds 
to taking $\phi =H_iK_i/c_i$ above, and then 
rescaling by the inverse of the  
product of the integrals, $1/H_i(\ep) K_i (\ep)$, 
in each case.
Thus the following statement holds.

\begin{theorem}

The maps (\ref{df56}) admit the compatible pair of  
invariant Poisson structures $(P^{(i)}(\ep),c_i^{-1}Q^{(i)}(\ep))$, where
$P^{(i)}(\ep)$ and $Q^{(i)}(\ep)$ are given respectively in Tables \ref{tab3} and  \ref{tab4}. The conformal factors $c_i$ are the same as 
in the continuous case, given in  Table \ref{tab1}.

\end{theorem}

\begin{table}[h!]\centering\caption{First deformed Poisson structure} \label{tab3}
\begin{tabular}{|l||c|c|c|}
 \hline
 $i$ & $P_{12}^{(i)}(\ep)$ & $P_{23}^{(i)}(\ep)$ & $P_{31}^{(i)}(\ep)$ \\ \hline \hline
 & & & \\
$1$ & $0$  &  ${\displaystyle y \left( 1 - \ep^2x^2\right)}$& $-x \left( 1 - \ep^2x^2\right)$ \\
 & & & \\
$2$ & $0$ & $ -y \left( 1 + \ep^2x^2\right)$ & $-x \left( 1 - \ep^2x^2\right) $\\
 & & &\\
$3$& $0$   & $y \left[ 1+\ep^2(x^2+y^2)\right]$ & $-x \left[ 1+\ep^2(x^2+y^2)\right]$\\
 & & & \\
$4$&  $0$ &  $-y \left[ 1-\ep^2(x^2-y^2) \right]$& $-x \left[ 1+\ep^2(x^2-y^2) \right]$\\
 & & &  \\
$5$& $0$ & $x \left( 1 - \ep^2y^2\right)$& $y \left( 1 + \ep^2x^2\right)$\\
 & & & \\
$6$ & $x \left( 1 + 3 \ep^2 y^2\right)$& $z \left( 1 + 3 \ep^2 y^2\right)$& 
$2 y \left( 1 - 3 \ep^2 x z\right)$\\
 & & &\\ \hline
\end{tabular}
\end{table}

\begin{table}[h!]\centering\caption{Second deformed Poisson structure} \label{tab4}
\begin{tabular}{|l||c|c|c|} \hline
$i$ & $Q_{12}^{(i)}(\ep)$ & $Q_{23}^{(i)}(\ep)$ & $Q_{31}^{(i)}(\ep)$  \\ \hline \hline
 & & &  \\
$1$ & 
$x $ & ${\displaystyle \frac{z+\ep^2 x (zx + 2 y^2)}{1-  \ep^2x^2 }}$ & $ 2 y$ \\
 & & &\\
$2$ & $x  \left( 1 - \ep^2x^2\right)$ & $z+ \ep^2 x (zx + 2  y^2)$ & $2 y \left( 1 -\ep^2x^2\right)$\\
 & & &\\
$3$
 & $z$  & ${\displaystyle \frac{x \left(1-\ep^2 z \right)}{1+\ep^2(x^2+y^2) }}$&
 ${\displaystyle \frac{y  \left(1-\ep^2 z \right)}{1+\ep^2(x^2+y^2) }}$\\
 & & & \\
$4$
 & $z  \left[ 1+\ep^2(x^2+y^2)\right]$& $x \left( 1 -\ep^2z^2\right)$& $y \left( 1 -\ep^2z^2\right)$ \\
 & & &  \\
$5$& 
$x \left( 1 + \ep^2x^2\right)$& $z -\ep^2 x (x z + 2y^2)$ & $2 y \left( 1 + \ep^2x^2\right)$\\
 & & & \\
$6$ &  $z + \frac32 \ep^2 x (x^2+y^2-z^2) $ & $x + \frac32 \ep^2 z (z^2+y^2-x^2) $& 
$y \left( 1 - 3 \ep^2 x z \right)$\\
 & & & \\ \hline
\end{tabular}
\end{table}

Note that Eqs. (\ref{p1}-\ref{q3}) provide one-parameter 
deformations of the Lie-Poisson tensors $P^{(i)}, Q^{(i)}$ given
in Table \ref{tab1}. 
This is equivalent to saying that  
Tables \ref{tab3} and  \ref{tab4} provide 
deformations of the real three-dimensional
Lie algebras 
$A_{3,3}, \mathfrak{sl}(2, \mathbb R), \mathfrak{so}(3), \mathfrak{e}(1,1), \mathfrak{e}(2)$.
Finally we 
note that the integrable discrete-time system 
$d \CE_6$ is just a particular case of the Hirota-Kimura discretization
of the $\mathfrak{so}(3)$ Euler top \cite{HK1}, 
whose bi-Hamiltonian structure has been presented recently in \cite{PS}.

\section{Explicit solutions to the integrable systems $d\CE_i$, $1 \leq i \leq 5$} \label{sec4}

As shown in \cite{HK1,HK2},
and recently in \cite{PS}, the integrable discrete-time systems obtained
through the Hirota-Kimura type discretization seem to admit a 
straightforward construction of their explicit solutions,
at least for the case of three-dimensional maps. 
Here we 
provide the explicit solutions for the discrete-time integrable systems $d\CE_i$ with
$1 \leq i \leq 5$. 
The cases $i=4,6$  
are each special 
cases of the $\mathfrak{so}(3)$ Euler top,
whose solutions, both continuous and discrete, are 
investigated in \cite{HK1,PS}, so here we present the solution 
only for $i=4$, since $i=6$ is similar. 
For comparison, 
in Table \ref{tab5} we give the explicit 
solutions for the continuous-time flows $\CE_i$ with $1 \leq i \leq 5$. The
parameters $\alpha,\beta,\gamma,\theta, \la ,\mu, k$  appearing in the 
table can easily be expressed in 
terms of the initial conditions and/or  
the integrals of motion.

{\footnotesize{
\begin{table}[h!]\centering\caption{Solutions to continuous systems $\CE_i$, 
$1 \leq i \leq 5$} \label{tab5}
\begin{tabular}{|l||c|c|c|| l|} \hline
$i$ & $x\left(t\right)$ & $y\left(t\right)$ & $z\left(t\right)$ & Integrals 
\\ \hline \hline
 & & &&  \\
$1$ & 
 ${\displaystyle \frac{1}{t+ \alpha}}$ & ${\displaystyle  \frac{\beta}{t+\alpha}}$ 
& ${\displaystyle  \gamma \left(t+\alpha\right) - \frac{\beta^2}{t+\alpha}}$ & $   \left\{ \begin{array}{l}
H_1 =\beta \\
K_1= \gamma
 \end{array} \right.$ \\
 & & &&\\
$2$ &${\displaystyle \frac{1}{t+ \alpha}}$  & $\beta \left(t+\alpha\right)$ & $\left(t+\alpha\right) 
\left[\gamma-\beta^2 \left(t+\alpha\right)^2 \right]$ & 
$   \left\{ \begin{array}{l}
H_2 =\beta \\
K_2= \gamma
 \end{array} \right.$ \\
 & & &&\\
$3$
 & $ {\displaystyle  
\frac{\la\cos\theta}{\cosh[ \la(t+\alpha)]} 
}$  & 
 $ {\displaystyle 
\frac{\la\sin\theta}{\cosh [\la(t+\alpha)]}
}$&
  $ {\displaystyle 
\la\tanh[ \la(t+\alpha)  ]
}$ & $
  \left\{ \begin{array}{l}
H_3 = \tan\theta  
\\
K_3 = \la^2/2 
 \end{array} \right.$ \\
 & & && \\ 
$4$& 
\small 
${\displaystyle 
\begin{array}{l} 
\frac{\la }{2}
\dnj\,[ \la (t+\alpha)] \\   
+ 
\frac{k\la }{2} 
\,\cnj\, [\la (t+\alpha) ]
\end{array} 
}$ & 
${\displaystyle 
\begin{array}{l} 
\frac{\la }{2}
\dnj \,[\la (t+\alpha)] \\
- 
\frac{k\la }{2} 
\,\cnj\,[\la (t+\alpha) ]
\end{array} 
}$ & 
$k\la\,\snj\,[ \la (t+\alpha)]$ &    
\normalsize 
$
   \left\{ \begin{array}{l}
H_4 =\la^2(1-k^2)/4  \\
K_4= \la^2(1+k^2) /4
 \end{array} \right.$ \\
 & & && \\ 
$5$& 
 $ {\displaystyle 
\la \, \mathrm{sech}\,[\la (t+\alpha)   ]
}$& $
 {\displaystyle 
-\la\tanh [  \la (t+\alpha)]
}$& 
 $ {\displaystyle 
\begin{array}{l}
\mu\cosh[\la (t+\alpha)] \\
+ \la\,  \mathrm{sech}\,[\la (t+\alpha)]
\end{array}
}$& $
 \left\{ \begin{array}{l}
  H_5 =\la^2/2   \\
  K_5 = \la (\mu+\la) 
  \end{array} \right.$ \\
 & & &  & \\ \hline
\end{tabular}
\end{table}
}}

We now construct the explicit solutions to the discrete-time systems $d\CE_i$ with $1 \leq i \leq 5$, thus
providing the discrete counterpart of Table \ref{tab5}. Let us recall that we consider each of $x,y,z$ as functions
on $\ep \mathbb{Z}$. To simplify the
notation we set 
$ x = x_{n},  y = y_{n},  z = z_{n}$, so that
$\tilde x = x_{n+1}, \tilde y = y_{n+1}, \tilde z = z_{n+1}$. 
For the sake of brevity, 
henceforth the discrete integrals of motion
$H_i (\ep)$ and $K_i (\ep)$ in Table \ref{tab2} 
will be denoted respectively by $\hat H_i$ and $\hat K_i$, $1 \leq i \leq 6$.

The following statement holds.

\begin{theorem}\label{solns} 
The explicit solutions to the integrable maps  
$d\CE_i$, $1 \leq i \leq 5$, given by Eq. (\ref{df56}), read:
\begin{itemize}
\item $i=1$:
\bea
&& x_n = \frac{1}{2\ep (n+ \tau )}, \label{s1}  \\
&& y_n =  \frac{\beta}{2\ep (n+\tau )}, \\
&& z_n = 2\gamma \ep \left(n+\tau\right) - 
\frac{\beta^2\ep^{-1}+ \gamma\ep}{2\left(n+\tau\right)}, \label{s3} 
\eea
with $\hat H_1 =\beta$, $\hat K_1 =\gamma$;
\item $i=2$:
\bea 
&&x_n = \frac{1}{2\ep (n+ \tau )}, \label{s4}  \\
&&y_n = 2\beta\ep \left[ (n+\tau ) -  
\frac{1} 
{4(n+\tau )}\right] , \\
&& z_n = \ep\left[ 2(n+\tau ) - \frac{1}{2(n+\tau )} \right]
\left[ \gamma -\beta^2\ep^2(4(n+\tau )^2-1)\right]
, \label{s6} 
\eea
with $\hat H_2 =\beta$, $\hat K_2 =\gamma$;
\item $i=3$:
\bea
&& x_n =  \frac{ 
\cos\theta\, \sinh\delta 
}
{\ep  
\cosh (2\delta n+\kappa )
}, 
\label{s7}  \\
&& y_n = \frac{ 
\sin\theta\, \sinh\delta 
}
{ \ep  
\cosh (2\delta n+\kappa )
}, \label{s8} \\
&& z_n =  \ep^{-1} \tanh\delta\tanh (2\delta n+\kappa )
,\label{s9} 
\eea
with
$$
\hat H_3 =\tan\theta 
, \qquad
 \hat K_3 = \frac{ \tanh^2 \delta}{2\ep^2}  
;
$$
\item $i=4$:
\bea
&& x_n = 
\frac{\snj \, \delta }{2\ep}  
\left[ \frac{\dnj (2n\delta +\kappa)}{\cnj \,\delta } + 
\frac{k\,\cnj (2n\delta +\kappa )}
{\dnj \,\delta } 
\right], \label{sde4x} \\ 
&& y_n = \frac{ \snj \, \delta  }{2\ep}
 \left[ \frac{\dnj (2n\delta+\kappa)}{\cnj\, \delta} -  
\frac{k \,\cnj (2n\delta+\kappa)}
{\dnj \,\delta} 
\right], \label{sde4y} \\ 
&& z_n =  \ep^{-1}\,  k\,  
\snj  \,\delta \, \snj (2n\delta+\kappa), \label{sde4z} 
\eea 
where $\snj $, $\cnj $, $\dnj $ are the Jacobian elliptic functions 
with modulus $k$ \cite{ww}, and 
\beq 
\label{de4hk} 
\hat H_4 = \frac{(1-k^2) \, \snj^2  \delta  }
{2\ep^2[2-(1+k^2)\snj^2  \delta ] }, \qquad 
\hat K_4 = \frac{1}{\ep^2}\left( 
\frac{1}{2}-\frac{\cnj^2  \delta  \, \dnj^2 \delta  }
{\cnj^2  \delta  + \dnj^2  \delta }
\right) 
; 
\eeq 
\item $i=5$:
\bea
&& x_n = 
\frac{  \sinh \delta } 
{\ep\cosh (2\delta n+\kappa)} 
, \label{s10} \\
&& y_n =  -\ep^{-1} \tanh \delta  \tanh (2 \delta n+\kappa) 
, \label{s11}  \\
&& z_n = \mu \cosh (2\delta n+\kappa) 
+(\ep^{-1}+\mu\sinh \delta)\sinh \delta\,\mathrm{sech}\, (2 \delta n+\kappa) 
,\label{s12} 
\eea
with
$$
\hat H_5 =\frac{1}{2\ep^2}\tanh^2 \delta 
, \qquad
 \hat K_5 =\ep^{-1}\mu\sinh \delta+\ep^{-2}\tanh^2 \delta.  
$$
\end{itemize}
\end{theorem}

{\bf Proof:} Let us illustrate the 
procedure to find the solutions (\ref{s1}-\ref{s12}) for just 
one of the five discrete systems
$d \CE_i$, $1 \leq i \leq 5$. We shall consider $i=3$.
The
remaining cases can be verified by elementary direct computations, apart from $d\CE_4$, which we reserve for the Appendix.

The system $d\CE_3$ reads:
\bea
&& x_{n+1}-x_n = -\ep (x_{n+1} z_n + x_n z_{n+1}), \label{m1} \\
&& y_{n+1}-y_n = - \ep (y_{n+1} z_n + y_n z_{n+1}), \label{m2}\\
&& z_{n+1}-z_n = 2\ep (x_{n+1} x_n + y_n y_{n+1}). \label{m3}
\eea
It has two integrals of motion,
$$
\hat H_3 = \frac{y_n}{x_n}=\frac{y_{n+1}}{x_{n+1}}, \qquad \hat K_3 = \frac 12 
\left[\frac{x_n^2 + y_n^2 +z_n^2}{1 + \ep^{2} (x_n^2 +y_n^2)}\right]=
 \frac 12 \left[\frac{x_{n+1}^2 + y_{n+1}^2 +z_{n+1}^2}{1 + \ep^2 (x_{n+1}^2 +y_{n+1}^2)}\right],
$$
in involution with respect to the pair $(P^{(3)} (\ep ), c_3^{-1}Q^{(3)}(\ep ))$, 
as given in Tables \ref{tab3} and \ref{tab4}.

The solution to the continuous-time flow $\CE_3$, as in Table \ref{tab5}, suggests the following ansatz for the solution of the map
(\ref{m1}-\ref{m3}):
\beq \label{t7}  
x_n = \frac{\nu\cos\theta}{\cosh T_n}, 
\qquad y_n =  
\frac{\nu\sin\theta}{\cosh T_n}, \qquad 
z_n = \la\tanh T_n ,
\eeq
with constant parameters $\la ,\nu,\theta$. 
By substituting the ansatz (\ref{t7}) 
into the formulae for the integrals, we see that $\hat H_3=\tan \theta$, 
while  
$$ 
\hat K_3= \frac{\la^2 +(\nu^2-\la^2)\mathrm{sech}^2T_n}
{2(1+\ep^2\nu^2\mathrm{sech}^2T_n)} 
$$ 
is constant (for all $T_n$) if and only if $\nu^2=\la^2/(1-\ep^2\la^2)$. 
Upon setting $\la = \ep^{-1}\tanh \delta$, in terms of another parameter 
$\delta$ (with $\delta /\ep =O(1)$ in the continuum limit $\ep\to 0$) 
this gives $\nu^2=\ep^{-2}\sinh^2\delta$ 
and $\hat K_3=\sinh^2\delta /(2\ep^2)$.      
Substituting the ansatz into the third part of the map, namely 
(\ref{m3}), and using the addition formulae for hyperbolic functions, 
one can 
see that this 
equation implies that 
$$\sinh (T_{n+1}-T_n)=\frac{2\sinh^2\delta}{\tanh\delta}
=\sinh 2\delta , $$  
hence $T_{n+1}-T_n=2\delta$. This implies 
that $T_n=2\delta n+\kappa$ for some constant 
$\kappa$, and then it is straightforward to verify 
that   
Eqs.  (\ref{m1}) and (\ref{m2}) 
are also satisfied identically. 
\endpf

\section{Discretization of three-dimensional bi-Hamiltonian flows 
with one transcendental invariant} \label{sec5}

There have been several studies of integrable Hamiltonian 
systems which have transcendental invariants \cite{giacomini, hietarinta}. 
Among the six bi-Hamiltonian flows with transcendental invariants 
listed in \cite{nutku} we select the following ones:
\bea
&&\CE_7: \qquad \dot x = -x^2, 
\; \; \, \; \; \; \; \, \; \; \; \;  \; \;  \; 
\dot y = -\xi x y, 
\qquad \qquad \; \; \;\; \;  \;
\dot z = 2 \xi y^2 +x z  \, ;\label{35} \\
&& \CE_8: \qquad \dot x = -x^2, 
\; \; \, \; \; \; \; \, \; \; \; \;  \; \;  \; 
\dot y = -x (x+y), 
\qquad \; \; \; \; \; \; 
\dot z = 2 y (x+y) +x z \,  ;\label{25} \\
&& \CE_9: \qquad \dot x = -xz, 
\; \; \, \; \; \; \; \, \; \; \; \;  \; \;  \,
\dot y = -\xi yz, 
\qquad \; \; \; \; \; \,  \; \qquad
\dot z = x^2 +\xi y^2 \,  .\label{sys5_9} 
\eea
In \cite{w} the real parameter $\xi$ is restricted to the range 
$|\xi| \in (0,1)$, but here we need not impose this 
requirement. 
Observe that the equations of motion (\ref{35}) 
reduce to the flow $\CE_1$ if $\xi=1$ and the flow 
$\CE_2$ if $\xi=-1$. Also, the equations (\ref{sys5_9}) reduce 
to $\CE_3$ if $\xi=1$, and to 
$\CE_4$ if $\xi=-1$. 

The Lenard-Magri chains for the flows (\ref{35}-\ref{sys5_9}) are given  by
\beq
 P^{(i)} d H_i=0, \qquad
P^{(i)} d K_i = \frac{1}{c_i} Q^{(i)} d H_i\qquad 
Q^{(i)} d K_i=0, \nonumber  
\eeq
for $i=7,8,9$ respectively, 
with $Q^{(7)}= Q^{(8)}= P^{(6)}$ and 
$Q^{(9)}=Q^{(6)}$ 
(related to $\mathfrak{sl}(2,\mathbb R)$ 
and to $\mathfrak{so}(3)$ respectively, see Table \ref{tab1}), 
\bea
&&P^{(7)}:\qquad P_{12}^{(7)} = \{x,y \}=0, \qquad P_{23}^{(7)} = \{y,z \}=\xi y, \qquad \; \; \; \; \, P_{31}^{(7)} = \{z,x \}=-x, \nonumber \\
&&P^{(8)}:\qquad P_{12}^{(8)} = \{x,y \}=0, \qquad P_{23}^{(8)} = \{y,z \}=x+y, \qquad  P_{31}^{(8)} = \{z,x \}=-x, \nonumber 
\eea
and $P^{(9)}=P^{(7)}$, with  
\bea
&& H_7 = y x^{-\xi}, \qquad \; \; \; K_7=H_6 = zx + y^2, \qquad  \qquad
\quad \;  \; \,
\;  \; \, c_7= x^{-(\xi+1)}, \nonumber \\
&& H_8 = x e^{-y/x}, \qquad 
K_8=H_6 = zx + y^2, \qquad  \qquad
\quad \;  \; \,
\; \; \; c_8= \frac{e^{-y/x}}{x}, \nonumber 
\\
&& H_9 = y x^{-\xi}, \qquad \; \; \; K_9=K_6 
=\frac{1}{2}(x^2+y^2+ z^2), \qquad  \;  \; \; \,  c_9=
x^{-(\xi+1)}. 
\nonumber 
\eea
Thus the transcendental invariants in each case 
are given by $H_7$, $H_8$ and $H_9$ respectively.  
(Strictly speaking, $H_7=H_9$ is only 
transcendental when $\xi\not\in\mathbb{Q}$, 
otherwise it is algebraic.) 
Moreover, note that the Lie algebra related to
$P^{(7)}$ is actually a one-parameter family of Lie algebras, 
parametrized by $\xi$; 
see 
\cite{w} for more details.
It can also be regarded as a four-dimensional Lie algebra, by 
taking $\hat{y}=\log y$ as a new coordinate and regarding 
$\xi$ as a central element. 

We now construct the Hirota-Kimura type discretizations of the flows 
$\CE_7$, $\CE_8$ and $\CE_9$; these are denoted 
using the notation introduced
in section \ref{sec4}.

\subsection{Explicit solutions to $d\CE_7$} 

The explicit solution to the equations of motion (\ref{35})  is given by:
\beq
x(t)=\frac{1}{t+\alpha}, \label{b1} \quad  
y(t)=\beta (t+\alpha)^{-\xi}, \quad   
z(t)=(t +\alpha)\left[\gamma - \beta^2(t+\alpha)^{-2\xi}\right], 
\eeq
with $H_7= \beta$ and $K_7 =\gamma$. 
Following the approach described in section \ref{sec4}, 
the discrete-time version of the flow $\CE_7$ reads:
\bea
&& x_{n+1}-x_n = - 2\ep x_{n}  x_{n+1}, \label{e1} \\
&& y_{n+1}-y_n = - \ep \xi  (y_{n+1} x_n + y_n x_{n+1}), \label{e2}\\
&& z_{n+1}-z_n = \ep (x_{n+1} z_n + x_n z_{n+1}) 
+ 4 \ep\xi  y_n y_{n+1}. \label{e3}
\eea
The decoupled equation for $x_n$ can be rewritten as a total difference, 
$$ 
\frac{1}{x_{n+1}}-\frac{1}{x_n}=2\ep , 
$$ 
from which it follows by summation that 
\beq
x_n= \frac{1}{2\ep (n+ \tau)}, \qquad \tau 
= \frac{1}{2\ep x_0} \label{ss};
\eeq
this is the discrete version of the first equation in (\ref{b1}), to which it tends in the continuum limit 
$$\ep\to 0, \qquad  2\ep n\to t, \qquad 2\ep\tau\to \alpha.$$ 
By substituting $x_n$ given by Eq. (\ref{ss}) into Eq.
(\ref{e2}) we get a difference equation for the variable $y_n$, whose solution reads
\beq
y_n= \frac{\tau \beta}{2\ep (n+\tau )} 
\frac{\Gamma \left(n+1+\tau - \xi/2 \right)  
\Gamma \left(\tau + \xi/2 \right)}
{\Gamma \left(n+\tau + \xi/2 \right) 
\Gamma \left(\tau + 1 - \xi/2 \right)}, \label{ww}
\eeq
where $\Gamma(z)$ is the complete gamma function.
We can now solve Eq. (\ref{ww}) 
for the constant $\beta$ (up to scale) to write it as a function of $x_n$ and 
$y_n$, which gives an explicit transcendental integral: 
$$ 
\hat H_7 =\frac {y_n}{x_n} 
\frac{\Gamma \left( \xi /2+(2\ep x_n)^{-1} \right)} 
{\Gamma \left(1-\xi /2+(2\ep x_n)^{-1} \right)}
. 
$$ 

Now inserting 
$x_n$ and $y_n$, given respectively by Eqs. (\ref{ss}-\ref{ww}) into Eq. (\ref{e3}) we find a difference
equation for $z_n$. Its solution is
\beq
z_n= \frac{\tau \gamma [4(n+\tau)^2-1]} 
{2\ep (4 \tau^2-1)(n+\tau)}+
\frac{2 \tau^2 \beta^2 \xi [4(n+\tau)^2-1] \Gamma^2 
\left(\tau +\xi/2 \right)}{ \ep  
 \Gamma^2 \left(\tau+1 -\xi/2 \right)(n+\tau)} W_n ,\label{ew}
\eeq
where
$$
W_n= \sum_{j=0}^{n-1} 
\frac{[ 2 (j+1+\tau) - \xi ] \Gamma^2 \left(j+1+\tau -\xi/2 \right)}
{[ 2 (j+\tau) +3 ] [ 2 (j+\tau) +\xi ] [ 4 (j+\tau)^2 -1 ] 
\Gamma^2 \left(j+\tau +\xi/2 \right) }.
$$
In principle, 
Eq. (\ref{ew}) can implicitly be solved for $\gamma$  
(after first replacing 
$j+\tau$ by $(2\ep x_j)^{-1}$ everywhere to remove 
explicit dependence on the parameter $\tau$), 
to give another transcendental invariant 
$\hat K_7$, 
in that 
case the bi-Hamiltonian structure can be reconstructed by the same formulae as above in 
cases 1--6; this means that the system $d\CE_7$ is completely 
integrable. Using the formula for $\hat{H}_7$ above we can 
reconstruct one invariant Poisson bracket for this map explicitly, as 
\bea
&& \{ x,y\}=0, \nonumber \\
&& \{ y,z\}=y(1-\ep^2 x^2)\left[
\frac{\Psi (1-\xi /2+ (2\ep x)^{-1})-\Psi (\xi /2+(2\ep x)^{-1})}{2\ep x}-1  
\right], \nonumber \\
&& \{ z,x\}=x(1-\ep^2 x^2) , \nonumber 
\eea
where $\Psi$ is the digamma function. This bracket has $\hat{H}_7$
as a Casimir, and for $\xi =\pm 1$ (up to scaling) 
it reduces to the brackets $P^{(1)}(\ep )$ and $P^{(2)}(\ep )$ respectively.  

It is straightforward to verify that the explicit form of the solution 
for $x_n,y_n,z_n$ given by Eqs. (\ref{ss}-\ref{ew})
can be used to recover  
the previous formulae for  
the discrete systems $d\CE_1$ and $d\CE_2$ given 
in Eqs. (\ref{s1}-\ref{s3}) and (\ref{s4}-\ref{s6}) by 
setting $\xi= \pm 1$  in the respective cases.  

\subsection{Explicit solutions to $d\CE_8$} 

The explicit solution 
to the equations of motion (\ref{25})  is given by 
\beq
x(t)=\frac{1}{t+\alpha}, \quad \label{b12} 
 y(t)=\frac{\beta- \ln (t+\alpha)}{t+\alpha}, \quad 
 z(t)=\gamma (t +\alpha)- \frac{ [\beta- \ln (t+\alpha)]^2}{t+\alpha},   \nonumber
\eeq
with $H_8= e^{-\beta}$ and $K_8 =\gamma$.
The discrete-time version of the flow $\CE_8$ reads:
\bea
&& x_{n+1}-x_n = - 2\ep x_{n}  x_{n+1}, \label{e12} \\
&& y_{n+1}-y_n = - \ep   (y_{n+1} x_n + y_n x_{n+1})  - 2\ep x_{n}  x_{n+1}, \label{e22}\\
&& z_{n+1}-z_n = \ep (x_{n+1} z_n + x_n z_{n+1} ) 
+2\ep (y_{n+1} x_n + y_n x_{n+1})+ 4\ep y_{n}  y_{n+1}. \label{e32}
\eea
The first equation for $x_n$ is identical to that in the previous case, 
and has the solution 
$ 
x_n= (n+ \tau)^{-1}/(2\ep ) 
$ 
as before. 
By substituting $x_n$ into Eq.
(\ref{e22}) we get a difference equation for the variable $y_n$, whose solution reads
\beq
y_n= \frac{\tau \beta - U_n}{2\ep (n+\tau)}, \label{ww2}
\eeq
where
$$
U_n = \Psi \left( n + \tau + 1/2 \right) - \Psi \left(\tau + 1/2 \right),
$$
with $\Psi(z)$ denoting the digamma function as before.
This leads to the transcendental invariant 
$$ 
\hat H_8 = \frac{y_n}{x_n} + \Psi \left(1/2 + (2\ep x_n)^{-1}\right). 
$$ 

Upon inserting $x_n$ as in (\ref{ss}) and $y_n$ given by Eq. (\ref{ww2}) 
into Eq. (\ref{e32}) we find a difference
equation for $z_n$, whose solution is given by
\beq 
\label{jbn}
z_n= \frac{\tau 
\left[  4 n (n+ 2 \tau) ( \beta^2 \tau + \beta +\gamma) 
+ 4 \tau^2 \gamma - \gamma\right]}
{ 2\ep \left(  4 \tau^2 -1\right)(n+ \tau)}-
\frac{2 \left[ 4(n+\tau)^2 -1 \right]}{2\ep (n+\tau)} V_n \,, 
\eeq
where
$$
V_n= \sum_{j=0}^{n-1} 
\frac{1 + 2 \beta \tau  
+ U_j [2(j+\tau)(1+2 \beta \tau)-1 + 2 \beta \tau]- U_j^2 [2(j+\tau)+1]}
{[ 2 (j+\tau) +3 ] [ 2 (j+\tau) +1 ] [ 4 (j+\tau)^2 -1 ] } .
$$
Similarly to the situation for 
$d\CE_7$, the system 
$d\CE_8$ has another transcendental integral $\hat K_8$ which is 
given implicitly  
by solving Eq. (\ref{jbn}) for $\gamma$. 
This implies that $d\CE_8$ is also bi-Hamiltonian and 
hence completely integrable.

\subsection{The system $d\CE_9$}

For all values of the parameter $\xi$, the equations of motion 
(\ref{sys5_9}) can be reduced to a quadrature, namely 
\beq \nonumber 
t+\mathrm{const} = \pm \int^{x(t)} \frac{ds}{s\sqrt{2H-s^2-K^2s^{2\xi}}}.  
\eeq 
Given $x(t)$ determined by this quadrature, $y$ and $z$ are then given by 
$$ 
y(t)=Kx(t)^{\xi}, \qquad 
z(t)=\pm\sqrt{2H-x(t)^2-K^2x(t)^{2\xi}} \,.
$$  
The constants $H$ and $K$ are respectively 
the values of $H_9$ and $K_9$ 
along an orbit. For certain values of $\xi$ the quadrature can be performed 
explicitly; for instance, when $\xi = 1$ it becomes an elementary integral, 
and the problem reduces to the solution of $\CE_3$, while for 
when $\xi = -1$ it becomes an elliptic integral, 
corresponding to the solution of $\CE_4$, as given 
in Table 5. The case $\xi =1/2$ is also an elementary one, 
while $\xi =-1/2$ and $\xi =2$ also give elliptic integrals  
(of the first and third kind, respectively). 
More generally, for all rational values of $\xi$ this quadrature 
is an hyperelliptic 
integral. 

However, it is straightforward to check that 
the cases $\xi =\pm 1$, which were solved already, are the only ones 
for which the system has the Painlev\'e property (i.e. all solutions 
are meromorphic functions of $t$ in these cases only). In general the 
solutions have movable algebraic branch points in the complex $t$ plane when 
$\xi \in \mathbb{Q}$, and movable logarithmic branch points  
when $\xi \not \in \mathbb{Q}$. 

The qualitative nature of the solutions is fairly insensitive to 
the parameter $\xi$. In fact, for $\xi>0$ the trajectories 
interpolate between the two fixed points $(x,y,z)=(0,0,\pm \sqrt{2H})$,  at  
the north/south poles of the sphere $x^2+y^2+z^2=2H$, 
while for $\xi<0$ there are closed periodic orbits. These two 
types of behaviour are exemplified by each of 
the explicitly solvable cases $\xi =\pm 1$.  
   
The Kahan-Hirota-Kimura discretization of 
this flow is given by 
\bea
&& x_{n+1}-x_n = - \ep (x_{n}  z_{n+1} + x_{n+1}  z_{n}), \nonumber \\
&& y_{n+1}-y_n = - \ep \xi (y_{n+1} z_n + y_n z_{n+1})  , \nonumber
\\
&& z_{n+1}-z_n = 2\ep x_{n+1} x_n  
+2\ep \xi y_{n+1} y_n.   
\nonumber
\eea
We have not attempted to solve this discrete system in the case   
$\xi \neq \pm 1$. In fact, numerical results for the latter  case 
(as described in the next section) provide evidence for the 
non-integrability of the system for generic values of $\xi$.  

\section{Diophantine integrability test} 

Over the past fifteen years or so there has been a 
gradual development of methods for testing integrability 
of maps or difference equations, using such  concepts as 
singularity confinement \cite{grp}, algebraic entropy  
\cite{hv}, Nevanlinna theory \cite{ahh} and orbit counting 
over finite fields \cite{rv}. In certain limited cases it has been 
proved that these tests provide necessary conditions for 
integrability of a map, in a suitable sense, 
most usually in the setting of 
algebraic integrability (see \cite{lg}, for instance), 
but in general it is an open 
problem to determine when these tests are effective. 

Most recently 
Halburd proposed an 
extremely simple criterion for integrability which applies 
to  rational maps 
defined over $\mathbb{Q}$ (or more generally over a number field), 
which he named the Diophantine integrability  
test \cite{halburd}.  
For a map whose $n$-th iterate has components $x_n\in \mathbb{Q}$, 
written as a fraction $x_n=p_n/q_n$ in lowest terms, the height of 
$x_n$   
is defined to be $H(x_n)=\mathrm{max}(\vert p_n\vert ,\vert q_n\vert )$; 
this is the archimidean height of $x_n$, and the logarithmic 
height is $h(x_n)=\log H(x_n)$. For a map in dimension $N$, with 
$N$ components, the height $H_n$ of the $n$-th point on an orbit is defined 
to be the maximum of the heights of all the components, with $h_n=\log H_n$ 
being  the logarithmic
height. 
Halburd defined a map to be \textit{Diophantine integrable} 
if the logarithmic height $h_n$ of the iterates of all orbits 
has at most polynomial growth in $n$. If we define 
the Diophantine entropy along an orbit $\mathcal{O}$ to be 
\[ 
E (\mathcal{O}):=\lim_{n\to\infty} \frac{1}{n}\log h_n,  
\] 
then a Diophantine integrable map is one for which 
$E (\mathcal{O})=0$ for all orbits.   

\begin{figure}[ht!]
\centerline{
\scalebox{0.3}{\includegraphics[angle=270]{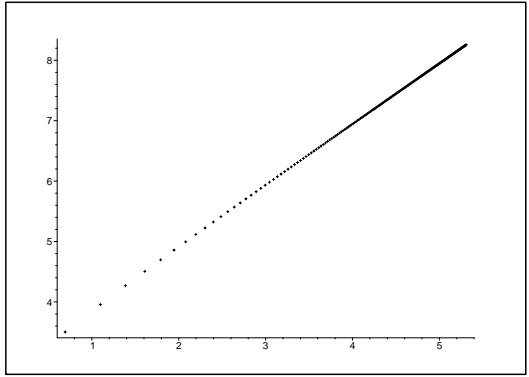}}
}
\caption{ {\it Plot of 
$\log h(x_n)$ versus $\log n$ 
for  
the first 200 iterates of 
the integrable map $d\CE_3$ for $\ep = 1/2$ with initial conditions 
$x_0=3/7,y_0=11/13,z_0=23/47$.  
} }
\label{dE3} 
\end{figure}

\begin{figure}[ht!]
\centerline{
\scalebox{0.3}{\includegraphics[angle=270]{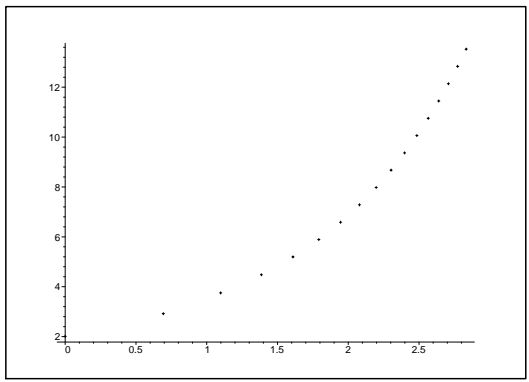}}
}
\caption{ {\it 
Plot of
$\log h(x_n)$ versus $\log n$
for
the first 17 iterates of Kahan's 
discretization of the Lotka-Volterra system  
with initial conditions
$x_0=1/2,y_0=5/3$ and parameter 
$\epsilon =9/14$. 
} }
\label{dlv}
\end{figure}

Diophantine entropy is somewhat similar to algebraic entropy \cite{hv}, which 
measures the height growth of rational functions generated by rational 
maps. In the latter setting the height of each iterate is 
just the maximum of the 
degrees of the polynomials in the numerator and denominator, considered 
as a rational 
function in the initial data. However,  
a huge disadvantage of using algebraic entropy 
is that one must usually try to guess a recursive relation 
to generate the degrees of these polynomials.       
The great advantage of Halburd's test is that it is extremely quick 
and straightforward to implemement numerically with a computer,  
and if the map is Diophantine integrable then a plot of 
$\log h_n$ against $\log n$ should look asymptotically like a straight line  
(see Figure \ref{dE3}), otherwise it will have an exponential shape 
(see Figure \ref{dlv}). 
The main drawback of using the test is that at present it has the 
status of a distinct definition of integrability, and it is 
not clear how it is related to other such definitions, like  
complete integrability in the Liouville-Arnold sense. 


Despite these drawbacks, it is worth remarking that, at least 
for maps in two or three dimensions, Diophantine integrability 
is a necessary condition for algebraic integrability. For example, 
a two-dimensional map which is algebraically integrable 
has a conserved 
quantity whose level sets are algebraic curves. Assuming that each of 
these curves 
is irreducible, and that not all orbits of the map are  
periodic, it was observed by Veselov \cite{ves}  
that they must all have genus zero or one; this follows from 
a theorem of Hurwitz which    
says that curves of genus two or more have automorphism groups of  
finite order \cite{Miranda}. (This argument 
also extends   
to the case when the level curves are reducible.)  
If the curve is rational (genus zero), then the map can be linearized, 
in which case the logarithmic heights grow linearly, $h_n\sim Cn$ for 
some constant $C$, while a curve of genus one is birationally 
equivalent to an 
elliptic curve, for which the heights grow as $h_n\sim Cn^2$.  
(See chapter 17 in \cite{cassels} for an introduction 
to archimidean heights on elliptic curves, or chapter VIII 
in \cite{silverman} for a more general discussion of heights.) 
Similar considerations apply to algebraically integrable  
maps in three dimensions, where the algebraic curves 
are the level sets of two independent integrals, or to systems 
with $N -1$ algebraic integrals in $N$ dimensions (as considered 
in \cite{lg} from the viewpoint of singularity confinement). 
However, in general these level sets can have two or more 
irreducible components; see \cite{hky} for several examples with 
two components in three dimensions.   

Here we prove that all of the discrete systems constructed 
here, except for $d\CE_9$, pass the Diophantine integrability test, before 
presenting numerical results which show 
more detailed behaviour of the growth of heights for 
some of these systems. For the theoretical and numerical analysis 
here it is convenient to set $\epsilon =1/2$; since the right hand 
sides of the difference equations are homogeneous (of degree two), 
this can always be achieved by scaling $x_n,y_n,z_n$ by the same factor. 
    
\begin{theorem} 
The discrete systems $d\CE_i$ for $i=1,\ldots ,8$ are all 
Diophantine integrable.  
\end{theorem}
{\bf Proof:} 
Without loss of generality 
we set $\epsilon =1/2$, as mentioned above, 
and consider each of the maps with rational initial data 
$x_0,y_0,z_0$ (and parameter $\xi\in\mathbb{Q}$ for the case of 
$d\CE_7$). This implies that all of the iterates $(x_n,y_n,z_n)$  
of these birational maps are also rational numbers for all $n$ 
(except on a set of initial data where these maps become singular). 
 
For the maps $d\CE_1$ and $d\CE_2$ it is clear from the explicit 
solutions, as  
given in Eqs. (\ref{s1}-\ref{s3}) and Eqs. (\ref{s4}-\ref{s6}) 
respectively, that in each case the iterates are given in terms of parameters 
$\alpha,\beta,\gamma\in\mathbb{Q}$, and these rational iterates 
have numerators and denominators which grow linearly in $n$. 
Hence the logarithmic height satisfies $h_n=\log n +O(1)$ (sub-polynomial 
growth) for these two maps. 

The maps $d\CE_3$ and $d\CE_5$ are naturally considered together, because 
their explicit solutions given in
Theorem \ref{solns} are in terms of hyperbolic functions 
(or equivalently, exponential functions of $n$) 
in
each case, which means that  
the intersections of the level sets of their two integrals are curves 
of genus zero. 
This implies that the heights of iterates should grow like 
$h_n\sim Cn$. To prove this directly for $d\CE_3$, note that one can 
eliminate $y_n$ from Eq. (\ref{m3}) by setting $y_n=\hat H_3 x_n$, and then 
further eliminate $z_n$ between that equation and Eq. (\ref{m1}) to 
get an expression of the form $z_n=F(x_n,x_{n+1},\hat H_3)$ with $F$ 
being a rational function.  
This leads to   
a single recurrence of second order for 
$w_n=1/x_n$, namely 
$$ 
w_{n+2}=\frac{4w_{n+1}^3+(1+\hat H_3^2)(2w_{n+1}+w_n)}
{4w_nw_{n+1}-(1+\hat H_3^2)}. 
$$
The latter recurrence has the conserved 
quantity 
$$ 
\hat L = \frac{2(w_n^2+w_{n+1}^2)+1+\hat H_3^2}{4w_nw_{n+1}-(1+\hat H_3^2)}, 
$$ 
and furthermore admits the linearization 
\beq\label{linde3} 
w_{n+2}-2\hat L w_{n+1}+w_n=0, 
\eeq 
which linearizes the system $d\CE_3$; 
in terms of the original integrals and solution parameters we 
find 
$ 
\hat L = (2+\hat K_3)/(2-\hat K_3)=\cosh 2\delta  
$.               
From the second order linear recurrence (\ref{linde3}) it follows 
directly that the height $H(w_n)$ grows exponentially with $n$, 
and hence 
$h(w_n)=h(x_n)\sim Cn$ (cf. Figure \ref{dE3}) for some 
$C>0$. Since $y_n=\hat H_3 x_n$, 
and $z_n$ can be written as a rational function of $x_n$ and $x_{n+1}$, 
it follows that $h(y_n)$ and $h(z_n)$ also have linear growth in $n$. 
Analogous arguments apply to $d\CE_5$. 
  

\begin{figure}[ht!]
\centerline{
\scalebox{0.3}{\includegraphics[angle=270]{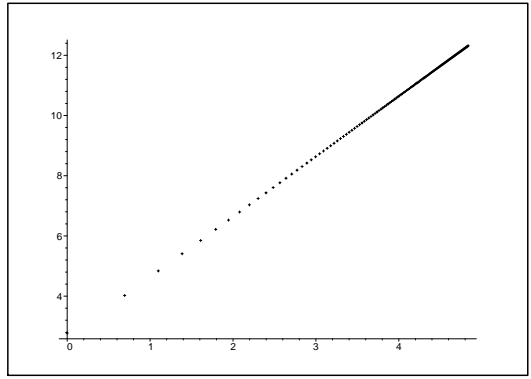}}
}
\caption{ {\it Plot of
$\log h(x_n)$ versus $\log n$
for 
the first 125 iterates of
the integrable map $d\CE_4$ for $\ep = 1/2$ with initial conditions
$x_0=7/3,y_0=11/13,z_0=23/47$.
} }
\label{dE4}
\end{figure}
Similarly, it is natural to 
consider the maps $d\CE_4$ and $d\CE_6$ together, because
the intersections of the level sets of their two integrals are curves
of genus one; the details for $d\CE_4$ are given in the Appendix.  
For $d\CE_4$  each of the coordinates $x_n,y_n,z_n$ of a point on an 
orbit can be written in terms of Jacobi functions, which are related 
by a M\"obius transformation to the Weierstrass $\wp$ function. 
For instance, the solution for $z_n$ in (\ref{sde4z}) 
is linear in the Jacobi sine, 
which is an elliptic function of order two 
with two simple poles in each period parallelogram; this implies 
that a relation of the form 
$
z_n = (aX_n+b)/(cX_n+d) 
$ 
holds, for some constants $a,b,c,d$, 
where $X_n$ is the $n$th term in 
a sequence of $X$ coordinates of points 
$P_0+nP\in \mathrm{E}$, for   
an elliptic 
curve $\mathrm{E}$ 
given in Weierstrass form as $Y^2=X^3+AX+B$ (for some $A,B$). 
It is known that, as long as $P$ is not a torsion point (which 
would correspond to a periodic orbit), the height grows like   
$h(X_n)\sim Cn^2$ as $n\to\infty$, where the constant $C>0$ 
only depends on the height of the point $P$ \cite{silverman}. 
Since $z_n$ is related to $X_n$ by a rational map of degree one, it 
follows that $h(z_n)$ has the same quadratic growth in $n$, and similarly 
for $h(x_n)$ and $h(y_n)$. The same arguments apply to $d\CE_6$, this being a 
special case of the Hirota-Kimura 
discrete Euler top, whose solutions are most 
naturally written in terms of Jacobi functions.  
           

\begin{figure}[ht!]
\centerline{
\scalebox{0.3}{\includegraphics[angle=270]{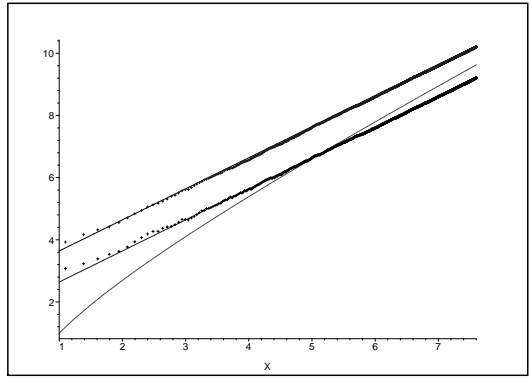}}
}
\caption{ {\it Plot of
$\log h(y_n)$ (bottom set of points) and 
$\log h(z_n)$ (top set of points)  versus $\log n$
for 
the first 2000 iterates of
the integrable map $d\CE_7$ for $\ep = 1/2$ with initial conditions
$x_0=3/7,y_0=11/13,z_0=23/47$ 
and parameter $\xi=19/17$. The bottom points have been fitted against 
$\log n +1.64$ (a straight line on this scale), and the top points 
against $\log n +2.64$; the curve $\log n +\log \log n$ is also 
shown.    
} }
\label{dE7}
\end{figure}
Finally, for the systems $d\CE_7$ and $d\CE_8$ we make use of 
direct estimates of the growth of heights, based on the original maps. 
For both these systems, note that from the explicit solution we 
have $h(x_n)=h(n+\tau)=\log n +O(1)$. It is convenient 
to define $Y_n=y_n/x_n$ and $Z_n=z_n/x_n$ in each case, and then  
note that $h(y_n)=h(Y_n)+O(\log n )$, and similarly for $h(z_n)$. 
From the second part of the map $d\CE_7$ we have 
\beq\label{yasy} 
Y_{n+1}=\left(\frac{n+\tau+1-\xi /2}{n+\tau+\xi /2}\right) 
\, Y_n,  
\eeq 
which implies 
$$
h(Y_{n+1})-h(Y_n)\leq \log n + O(1) \implies 
h(Y_n)\leq n\log n + O(n), 
$$ 
where the second implication follows by summing over $n$. 
Thus $h(Y_n)$ has weaker than quadratic growth in $n$. 
Similarly for $Z_n$ we have 
$$ 
(n+\tau-1/2)Z_{n+1}=(n+\tau+3/2)Z_{n}+2\xi Y_nY_{n+1}, 
$$ 
which implies that 
$$ 
h(Z_{n+1})\leq h(Z_n)+h(Y_nY_{n+1})+O(\log n)\leq 
h(Z_n)+n\log n+ O(n),
$$ 
and hence 
$ 
h(Z_n)\leq\frac{1}{2}n^2\log n + O(n^2)    
$, 
which is weaker than cubic in $n$. For $d\CE_8$, analogous 
estimates show that 
$h(Y_n)\leq n\log n + O(n)$ and 
$h(Z_n)\leq 2 n^2\log n + O(n^2)$, so this system is 
Diophantine integrable as well. 
\endpf 
Having proved that the systems are all Diophantine integrable, we can 
compare the theoretical results with some numerical experiments. 
For the system $d\CE_3$ we see that the log-log plot gives 
what we expect: genus zero means 
linear growth of logarithmic height, so 
$\log h(x_n)= \log n + O(1)$; this is evident from the plot of points  
in Figure \ref{dE3}, which lie  
asymptotically on a straight
line of slope 1. Similarly for the genus one case, we expect  
$\log h(x_n)= 2\log n + O(1)$, and Figure \ref{dE3} shows 
points which asymptote to a line with slope 2. In this case the offset, 
corresponding to the correction at $O(1)$, is function of the 
height of a point on an associated 
elliptic curve, and both the point and the 
curve vary with the initial data of the map.   

\begin{figure}[ht!]
\centerline{
\scalebox{0.3}{\includegraphics[angle=270]{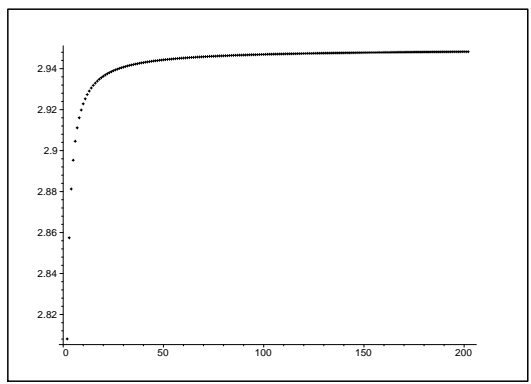}}
}
\caption{ {\it Plot of
$\log (h(x_n)/n)$ versus $n$
for $d\CE_3$ with the same data as Figure \ref{dE3}.  
} }
\label{de3correx}
\end{figure}

\begin{figure}[ht!]
\centerline{
\scalebox{0.3}{\includegraphics[angle=270]{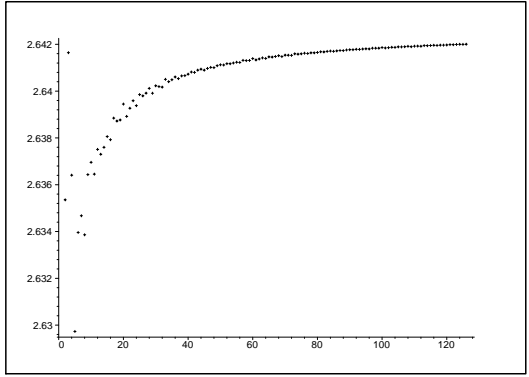}}
}
\caption{ {\it Plot of
$\log (h(x_n)/n^2)$ versus $n$
for 
$d\CE_4$ with 
the same data as Figure \ref{dE4}. 
} }
\label{de4correx}
\end{figure}

The theoretical results on the growth of heights for the 
algebraically integrable systems 
$d\CE_i$ for $1 \leq i \leq 6$, 
as detailed in the above proof, 
are confirmed by the numerical calculations, and for those cases we have 
an exact expression for the leading order asymptotic behaviour. Moreover, 
one can also look at how the asymptote is approached. 
Taking the system $d\CE_3$ 
for example, $h(x_n)/n$ approaches a 
constant as $n\to\infty$,  and from the numerical plot in 
Figure \ref{de3correx}
one can see that this limit is reached in a very uniform manner, 
in keeping with a correction of $O(1/n)$ to this constant. Similarly, 
in the case of $d\CE_4$,    
corresponding to motion on 
an elliptic  curve, we see from Figure \ref{de4correx}
that once again the convergence of $h(x_n)/n^2$ to a constant appears to 
be almost monotone.


The non-algebraically integrable cases, $d\CE_7$ and $d\CE_8$, have some 
extremely interesting features compared with the others. First of all, the 
method of proof used in Theorem 4 above has not necessarily 
provided the leading order 
asymptotics of the logarithmic heights, but has merely given 
upper 
bounds on the growth of the form $Cn^j \log n$ with $j=1$ for 
$h(y_n)$ and $j=2$ for $h(z_n)$ in each case.    
Let us focus on the case of $d\CE_7$. Upon looking more closely at 
Eq. (\ref{yasy}), it would appear that the upper bound for 
$h(Y_n)$ might be sharp, so that $h(Y_n)\sim n\log n$ (and  
$h(y_n)$ would have the same leading order asymptotics). However, 
studies of particular sequences of 
rational  iterates show that cancellations occur between the 
numerator and denominator of $Y_n$ and the prefactor 
$(n+\alpha+1-\xi /2)/(n+\alpha+\xi /2)$, which means that the 
height of $Y_{n+1}$ is therefore smaller than the crudest 
estimate for the upper bound. This weaker growth has a knock-on effect, 
meaning that 
the growth of heights of $z_n$ also seems to be much weaker than 
expected.  Indeed, Figure \ref{dE7} suggests 
that the correct asymptotics should be linear growth in $n$ for 
the logarithmic heights of both $y_n$ and $z_n$, i.e.  
$h(y_n)\sim C_1 n $  and $h(z_n)\sim C_2 n $ for positive constants 
$C_1,C_2$. For the particular sequence of heights plotted in 
that figure, a numerical fit shows that $\log C_1\approx 1.64$ 
and $\log C_2\approx 2.64\approx \log C_1+1$. We have also plotted 
$\log n+ \log \log n$ for comparison, to show how 
the upper bound for $h(y_n)$ fails to be sharp. Another surprising feature 
of this system is that for different choices of initial data 
we find (to within numerical accuracy) 
the same values of $C_1$ and $C_2$; this 
is in contrast to the algebraically integrable setting described 
above, where the coefficient in front of the leading order 
term is dependent on the initial data. Thus we might conjecture that 
for this map $C_1,C_2$ are independent of initial data, 
and also that $C_2=e\, C_1$ holds identically, in which case 
there should be some deeper 
arithmetical explanation for this asymptotic behaviour.   
Similarly to the case of $d\CE_7$, 
numerical results for the system $d\CE_8$ also show 
linear growth of logarithmic heights.  


\begin{figure}[ht!]
\centerline{
\scalebox{0.3}{\includegraphics[angle=270]{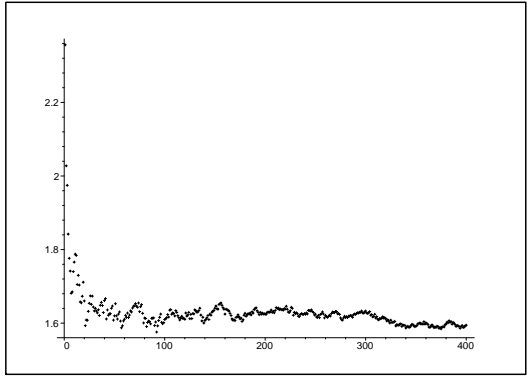}}
}
\caption{ {\it Plot of 400 iterates of   
$\log (h(x_n)/n)$ versus $n$
for
$d\CE_7$ with
the same initial data and parameters as in Figure \ref{dE7}.
} }
\label{de7correx}
\end{figure}
Supposing that the numerical observation of 
linear growth of $h_n$ for these discrete 
systems with transcendental invariants is indeed the correct 
asymptotic behaviour, it is then interesting to look at how 
$h_n/n$ approaches a constant value. The results we find are in stark 
contrast to the algebraic setting: rather than the 
almost monotone convergence seen in the previous examples, for 
$d\CE_7$ we find that $h_n/n$ shows rapid fluctuations which persist 
for increasing values of $n$. These fluctuations in the asymptotics are 
somewhat reminiscent of the ``random''-looking error terms that appear in 
some famous arithmetical functions, such as the difference between 
the prime-counting function $\pi (n)$ and the logarithmic integral 
\cite{mazur}. It would be interesting to know whether these 
fluctuations might provide a means of characterizing the 
difference between discrete systems which are algebraically integrable and 
those with transcendental invariants.


For comparison with the Diophantine integrable examples above, in 
Figure \ref{dlv}   
above we have plotted the growth of $h_n$ for a particular case of 
the discrete 
Lotka-Volterra system 
due to Kahan, which is the degree two birational map 
given in Eq. (\ref{kahandlv}). 
This figure shows that the logarithmic 
height seems to grow exponentially, 
indicating non-integrability of this system.  Indeed, 
the heights of 
iterates  
grow so fast that even on a fairly new computer it took 1 hour 
to calculate the heights of 17 rational iterates with Maple; the 
value of $h(x_{17})$ is of the order of $10^{325009}$ in this case. 
Upon examining the data used in Figure \ref{dlv} 
more carefully, it is apparent that 
$h(x_{n+1})\approx 2\, h(x_n)$ to very 
good accuracy, so we expect  $h_n\sim C\, 2^n$. This would mean that 
the logarithmic height essentially doubles with each step, giving  
a Diophantine entropy of $\log 2$ for generic (aperiodic) orbits. 
This appears to be the same as the 
algebraic entropy of the map when 
$\epsilon^2\neq 1$, which was calculated by A. Ramani \cite{ramaniprivate}. 
In any case, 
this is the entropy value that one would expect for 
a generic (non-integrable) birational map of degree two. 

\begin{figure}[ht!]
\centerline{
\scalebox{0.3}{\includegraphics[angle=270]{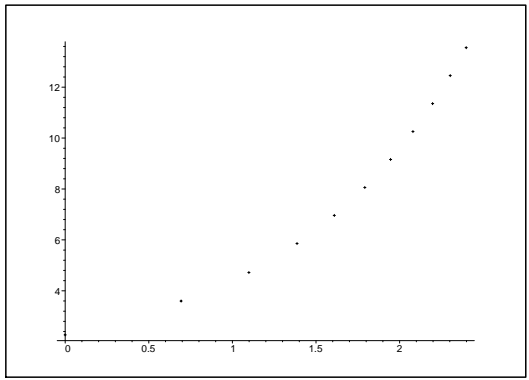}}
}
\caption{ {\it Plot of
$\log h(x_n)$ versus $\log n$
for 
the first 11 iterates of
the map $d\CE_9$ for $\ep = 1/2$ with initial conditions
$x_0=2/5,y_0=7/3,z_0=11/13$ and parameter $\xi =-1/2$.
} }
\label{dE9}
\end{figure}

Finally we should  mention the results of numerical calculations of the growth of heights 
for the map $d\CE_9$  
for various cases with $\xi\neq\pm 1$ 
(that is, excluding the two special cases where the map is already known to be algebraically 
integrable). For generic rational values of the parameter $\xi$ we find that the map  
$d\CE_9$ is not Diophantine integrable, but rather the 
Diophantine entropy is $\log 3$ for generic orbits, this being the typical 
value to be expected for a non-integrable birational map of degree three. 
(See Figure \ref{dE9} for an illustration of the exponential growth 
of logarithmc heights when $\xi=-1/2$.) 
These numerical results suggest that while 
that the original continuous system   
$\CE_9$ is algebraically integrable (in the sense of having 
two independent algebraic integrals), the corresponding discrete system  
is not. We shall return to this point in our conclusions. 
   

\section{Concluding remarks} \label{sec6}

In this paper we studied three-dimensional 
birational maps  which provide 
integrable time-discretizations
of quadratic 
bi-Hamiltonian flows associated with pairs of real three-dimensional Lie algebras,
as presented in \cite{bw,nutku}. We have shown that for the six cases  
of continuous flows which are 
algebraically integrable, 
the Hirota-Kimura  
type discretization provides 
maps 
admitting two independent rational integrals of motion, 
in involution
with respect to a pair of compatible
Poisson tensors. We have also provided 
explicit solutions of the resulting discrete systems 
$d\CE_i$ for $i=1,\ldots ,6$,  
which 
are given in terms of either rational, hyperbolic or elliptic functions 
in each case. 
These results confirm the conjecture 
that the property of algebraic integrability is preserved by 
this discretization scheme. 

We have also applied the same 
procedure to 
two cases of   
integrable continuous flows in three dimensions having one rational and one  
transcendental integral of motion, for which the  
resulting maps, $d\CE_7$ and $d\CE_8$, 
admit explicit solutions in terms of rational functions and either 
gamma or digamma functions. 
In each of these cases, we have 
found an explicit formula for one transcendental integral of the map, 
but the second integral is only defined implicitly by the solution. 
Nevertheless, this is sufficient to assert 
that the latter two cases are also 
completely  
integrable in the Liouville-Arnold sense. 
Therefore the Kahan-Hirota-Kimura discretization scheme preserves 
integrability even in these transcendental cases. 
However, for another example of a continuous integrable system 
with one  transcendental integral, it appears likely that the corresponding 
discretization $d\CE_9$ is not integrable for generic values of the 
parameter in the map. 
    
In an attempt to gain a better understanding of the 
difference between the algebraic and transcendental cases, 
we have also analyzed all of these  
discrete systems from a different viewpoint, within the arithmetical 
setting of 
Diophantine integrability.  
So far, the Diophantine integrability test has been 
applied to various algebraically integrable systems and discrete 
Painlev\'e equations \cite{halburd}, as well 
as to certain birational maps that are not algebraically integrable 
and fail the test \cite{hone, hone2}. 
Our theoretical results show that $d\CE_7$ and 
$d\CE_8$ provide examples of 
discrete integrable systems with transcendental invariants that 
are also Diophantine integrable, in the sense defined by Halburd.  
Moreover, more detailed numerical results suggest that these  
discrete integrable systems might be distinguished 
from algebraically integrable maps by the manner in which 
the logarithmic heights converge to their leading order asymptotics.    
This asymptotic behaviour deserves to be studied more carefully in 
the future. 

A further interesting point is 
the comparison with Kahan's 
discretization of the Lotka-Volterra system, which is an 
integrable flow in the plane with a transcendental integral.  
The discrete system 
provides a non-standard symplectic integrator of this flow, 
and seems to preserve the qualitative features of the continuous  
counterpart. 
However, the numerical results indicate that the discrete 
system is not Diophantine integrable, which adds further evidence 
to the conjecture that (for generic values of $\ep$)  
it should not be Liouville integrable either.  

Similar considerations 
apply to the discrete system $d\CE_9$: numerically it appears that for generic values of the 
parameter $\xi$, it fails the Diophantine integrability test. 
Since the continuous system $\CE_9$ has algebraic integrals for $\xi\in\mathbb{Q}$, 
this means that in general 
algebraic integrability (in the weakest sense of the term) is not preserved 
by the Kahan-Hirota-Kimura discretization. Actually one could already observe this 
for the system $\CE_7$, since it has the integral $H_7$ which 
is algebraic for all $\xi\in\mathbb{Q}$, 
but the integral $\hat H_7$ for 
$d\CE_7$ is transcendental unless $\xi\in\mathbb{Z}$. This suggests 
that one should impose a much stronger notion of algebraic complete integrability (a.c.i.) if 
this is to be preserved by the discretization scheme. For instance, one can require 
that the generic level sets of the integrals are smooth abelian varieties, possibly extended 
by $(\mathbb{C}^*)^m$ for some $m$. An excellent discussion of various 
different definitions of a.c.i. can be found in chapter V of 
\cite{vanhaecke}. 

As for the case of the discrete 
three-dimensional Euler top (see \cite{HK1,PS}),
there is one other standard 
attribute of integrable systems that remains to be
found for the maps $d\CE_i$ for $i=1,\ldots ,8$, 
namely their Lax representation. 
This is an open problem which deserves further investigation.

There are three other bi-Hamiltonian flows in the list of G\"umral 
and Nutku, all of which have one transcendental invariant. Preliminary 
results suggest that their Kahan-Hirota-Kimura type discretizations 
are qualitatively similar to the system  $d\CE_9$, and we expect that 
these maps are not Liouville integrable. We reserve the study of these systems 
for future work.

Finally we remark 
that another non-standard symplectic integrator for the Lotka-Volterra
model, with similar numerical properties, has been 
given by Mickens in \cite{mickens}. It would be very interesting to 
see if the approach to discretization proposed by Mickens 
shares some of the remarkable properties of Kahan's.

\section*{Acknowledgments}

AH and MP are grateful to the organisers for supporting their
attendance at the Miniworkshop on Integrable 
Systems at Universit\`a di Milano-Bicocca in September 2007, 
where this collaboration began. 
Both authors 
are grateful to Yuri Suris for helpful correspondence.  
AH would also like to thank Kim Towler for a careful reading of 
the text, and  
Gavin Brown for useful discussions.
 
\section*{Appendix: solution of the system $d\CE_4$ in elliptic functions} 

\begin{figure}[ht!]
\centerline{
\scalebox{0.3}{\includegraphics[angle=270]{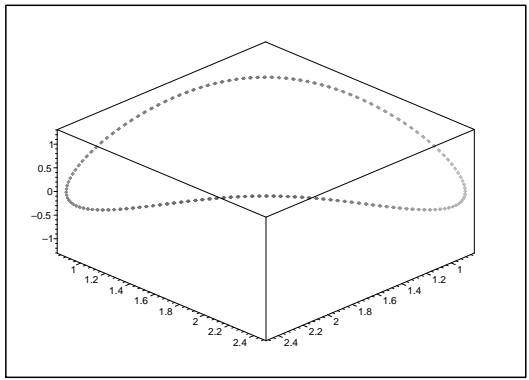}}
}
\caption{ {\it Plot of the first 1000 points on  
the orbit 
of the integrable map $d\CE_4$ with initial conditions
$x_0=7/3,y_0=11/13,z_0=23/47$.
} }
\label{dE4curve}
\end{figure}

Here we derive the formulae (\ref{sde4x}-\ref{sde4z}) corresponding to the solution of $d\CE_4$, 
as in Theorem \ref{solns}. The first observation to make is that the curve $\CV$ in affine three-dimensional 
space defined by the 
equations 
$$
\CV: \qquad xy=H\left[1+\ep^2(x^2+y^2)\right], \qquad 
x^2+y^2+z^2 = 2K \left[1+ \ep^2(x^2+y^2)\right], 
$$   
corresponding to the intersection of the level sets $\hat H_4=H$,  
$\hat K_4=K$, has genus one (at least for generic values of $H$ and $K$). 
To see this, note that $\CV$ is a double cover 
of the curve 
$$
\CC: \qquad xy=H\left[1+\ep^2(x^2+y^2)\right]
$$ 
in two dimensions, via the covering 
map 
$$
\begin{array}{lccc} 
\pi :  & \CV       & \rightarrow & \CC \\ 
       & (x,y,z) & \mapsto         & (x,y) 
\end{array} 
$$
which is ramified over the four points $(x,y)\in \CC$
obtained from  the simultaneous solutions of 
$xy=H(1+\ep^2(x^2+y^2))$, $(1-2K\ep^2)(x^2+y^2)=2K$ (when $z=0$). 
Since the curve $\CC$ is a conic (genus zero), it follows from the 
Riemann-Hurwitz formula \cite{Miranda} that  
$\CV$ is (the affine part of) a curve of genus one. 
The first order recurrence relations for $x_n,y_n,z_n$, namely
\bea
\label{de4eq} 
&& x_{n+1}-x_n = -\ep (x_{n+1}z_n + x_{n}  z_{n+1} ), \label{de4eq1}  \\
 &&  y_{n+1}-y_n = \ep  (y_{n+1} z_n + y_n z_{n+1})  
,   \label{de4eq2} \\
&& z_{n+1}-z_n = 2\ep (x_{n+1} x_n  - y_n y_{n+1}) ,  \label{de4eq3} 
\eea
correspond to a birational map from this curve to itself, 
inducing an automorphism of an isomorphic elliptic curve 
(i.e. a plane curve defined by a Weierstrass cubic), and it
follows that $x_n = X(u+nv)$ for a suitable elliptic function 
$X$, and similarly for $y_n,z_n$. 
One can see some points on a real 
connected component of such a curve in Figure \ref{dE4curve}. 

From the equations for $\CV$ it 
is easy to see that the functions corresponding to $x_n,y_n,z_n$ 
each have simple poles at the same places, and they are elliptic functions  
of order two. These facts suggest that it may be most convenient to write the formulae 
in terms of Jacobian (rather than Weierstrassian) elliptic functions. Indeed, 
if we set 
$$
s = x+y,\qquad d=x-y,
$$ 
then the equations for $\CV$ become 
$$ 
(1-2H\ep^2)s^2-(1+2H\ep^2)d^2=4H , \qquad (1-2K\ep^2)(s^2+d^2)+4z^2=4K ,
$$
which are reminiscent of (linear combinations of) 
the quadratic relations 
$$ 
\snj^2 (u) + \cnj^2 (u) =1, \qquad k^2\snj^2 (u) + \dnj^2 (u) =1, 
$$ 
for Jacobi functions. 

In order to obtain the formulae (\ref{sde4x}-\ref{sde4z}), it is instructive to take a detour through Jacobi 
theta functions, by deriving bilinear equations from Eqs. (\ref{de4eq1}-\ref{de4eq3}). Upon setting 
$$
x_n = \frac{A_n+B_n}{2\ep D_n}, \qquad
y_n = \frac{A_n-B_n}{2\ep D_n}, \qquad
z_n = \frac{C_n}{\ep D_n}, 
$$ 
the system (\ref{de4eq}) is equivalent to the following three bilinear equations: 
\begin{equation}\label{de4bil}
\renewcommand{\arraystretch}{1.3}
\left\{\begin{array}{l}
 A_{n+1}D_n-A_nD_{n+1} = - (B_{n+1}C_n + B_{n}  C_{n+1} ),\\
B_{n+1}D_n-B_nD_{n+1} = -(A_{n+1} C_n + A_n C_{n+1}),\\
C_{n+1}D_n-C_nD_{n+1} = A_{n+1} B_n  + A_n B_{n+1}.
\end{array}\right.
\end{equation}
One should hesitate to call (\ref{de4bil}) the Hirota bilinear form 
of Eqs. (\ref{de4eq1}-\ref{de4eq3}), because there are four unknowns (tau-functions) 
$A_n,B_n,C_n,D_n$ but 
only three equations, so the system is underdetermined. 
Despite this apparent problem, we can solve this bilinear 
system in terms of Jacobi theta functions $\vartheta_j$, $j=1,\ldots,4$, by comparing these equations with 
the identities in exercise number 3 on page 488 of \cite{ww}; the first of these 
is the relation 
\beq \label{thetarel}
\vartheta_1 (u\pm v) \vartheta_2 (u\mp v)\vartheta_3\vartheta_4 = 
\vartheta_1 (u) \vartheta_2 (u)\vartheta_3(v)\vartheta_4(v)\pm 
\vartheta_3 (u) \vartheta_4 (u)\vartheta_1(v)\vartheta_2(v)\,
\eeq 
the last is 
\beq \label{thetarel2}
\vartheta_3 (u\pm v) \vartheta_4 (u\mp v)\vartheta_3\vartheta_4 = 
\vartheta_3 (u) \vartheta_4 (u)\vartheta_3(v)\vartheta_4(v)\mp 
\vartheta_1 (u) \vartheta_2 (u)\vartheta_1(v)\vartheta_2(v)\,
\eeq 
and there are four other relations of this kind, for different permutations of the four indices.  
Here $\vartheta_j$ without argument denotes a theta constant (i.e. 
$\vartheta_j=\vartheta_j(0)$, which depends on the modulus $k$).
By taking the sum of the two equations given in (\ref{thetarel}) with 
opposite choices of $\pm$ signs, and similarly taking the difference of the two 
equations specified by (\ref{thetarel2}), one sees that both 
$\vartheta_1 (u+ v) \vartheta_2 (u- v) +\vartheta_1 (u- v) \vartheta_2 (u+ v)$ 
and 
$\vartheta_3 (u+ v) \vartheta_4 (u- v) -\vartheta_3 (u- v) \vartheta_4 (u+ v)$
are proportional to  $\vartheta_1 (u) \vartheta_2 (u)$, modulo $v$-dependent factors. 
Thus if $v$ is regarded as a fixed constant, and the shift $u\to u+2v$ is 
identified with $n\to n+1$, then the first equation 
in (\ref{de4bil}) is satisfied if (suppressing all arguments 
and the index $n$) the identifications 
$$ 
C\sim\vartheta_1, \quad  
B\sim\vartheta_2, \quad  A\sim\vartheta_3, \quad  D\sim\vartheta_4 
$$ 
are made, up to suitable $v$-dependent  scaling denoted by the $\sim$ symbol. 
Moreover, these identifications are consistent with the second and third 
equations in  (\ref{de4bil}), which are consequences of the aforementioned other 
four bilinear relations between Jacobi theta functions. 

Given that the Jacobian elliptic functions are defined in terms of theta functions by 
$$ 
\snj (u) = \frac{\vartheta_3}{\vartheta_2}\frac{\vartheta_1(u/\vartheta_3^2)}{\vartheta_4(u/\vartheta_3^2)},
\qquad 
\cnj (u) = \frac{\vartheta_4}{\vartheta_2}\frac{\vartheta_3(u/\vartheta_3^2)}{\vartheta_4(u/\vartheta_3^2)},
\qquad 
\dnj (u) = \frac{\vartheta_4}{\vartheta_3}\frac{\vartheta_3(u/\vartheta_3^2)}{\vartheta_4(u/\vartheta_3^2)},
\qquad 
$$ 
 it 
follows that the solution of the difference equations (\ref{de4eq}) has the 
form 
\bea
&&x_n = \la \, \dnj (\kappa+2n\delta) + \mu \,\cnj (\kappa+2n\delta), \nonumber \\ 
&&y_n = \la \,\dnj (\kappa+2n\delta) - \mu \,\cnj (\kappa+2n\delta), \nonumber \\ 
&&z_n = \nu \,\snj (\kappa+2n\delta),  \nonumber 
\eea
for constants $\delta , \kappa$ and suitable  
prefactors $\la , \mu , \nu$ which are given in terms of $\delta$ and the theta constants. 
The 
expressions (\ref{sde4x}-\ref{sde4z}) can also be verified directly from the 
addition formula for $\snj$, namely 
\beq 
 \label{snadd}
\snj (u+v) = 
\frac{\snj (u) \cnj (v)\dnj (v) + \snj (v) \cnj (u)\dnj (u) }{1-k^2\snj^2(u)\snj^2(v)},
\eeq 
as well as the analogous formulae for $\cnj$ and $\dnj$. 
Using (\ref{snadd}) to calculate $\snj (u+v) - \snj (u-v)$, 
and then setting $u\to \kappa +(2n+1)\delta $, $v \to \delta$, gives an expression 
for the left hand side of the third equation in (\ref{de4eq}), and performing analogous 
computations for the 
right hand side and for the other two difference equations allows the 
prefactors 
$\la , \mu , \nu$ to be determined directly in terms of Jacobi functions 
with argument $\delta$, in agreement with (\ref{sde4x}-\ref{sde4z}). 

Finally, note that the solution depends on the required number of 
arbitrary constants, namely the three parameters $\delta ,\kappa ,k$. The 
parameter $\delta$ and the modulus $k$ are determined by the values of the 
integrals $\hat H_4$ and $\hat K_4$, by solving the relations (\ref{de4hk})
as  a system for $k$ and $\snj \delta$ and then performing the elliptic 
integral $\delta =\int_0^{\snj \delta}d\xi /\eta$, while $\kappa$ is found from 
$$
\kappa=\int_0^{\frac{\ep z_0}{k\snj\delta}}\frac{d\xi}{\eta}, 
\qquad \eta^2=(1-\xi^2)(1-k^2\xi^2).
$$ 
 

\end{document}